\documentclass{iopjournal}

\usepackage{amsmath}
\usepackage{amssymb}
\usepackage{float}      
\usepackage{multirow}   
\usepackage[numbers,sort&compress]{natbib} 
\usepackage{orcidlink}

\begin{document}

\articletype{Paper} 

\title{Physics-Informed Neural Networks vs U-Net: Application to thermal dust emission recovery in the microwave}

\author{Nicolás M. Juan$^{1,2,*}$\orcid{0009-0005-1494-1981}, José M. Casas$^{2,3}$\orcid{0000-0001-7882-1691} and Laura Bonavera$^{3,4}$\orcid{0000-0001-8039-3876}}

\affil{$^1$Universidad Internacional de Valencia (VIU), C/Pintor Sorolla 21, 46002, Valencia, Spain}

\affil{$^2$Universidad de Oviedo, Departamento de Matem\'{a}ticas, C. Federico Garc\'{i}a Lorca 18, 33007 Oviedo, Spain}

\affil{$^3$Instituto Universitario de Ciencias y Tecnolog\'{i}as Espaciales de Asturias (ICTEA), C. Independencia 13, 33004 Oviedo, Spain}

\affil{$^4$Universidad de Oviedo, Departamento de F\'{i}sica, C. Federico Garc\'{i}a Lorca 18, 33007 Oviedo, Spain}

\email{nicomorantjuan@gmail.com}

\keywords{Inverse Problems, Deep Learning, Interstellar Medium}

\begin{abstract}
Galactic thermal dust emission is the dominant foreground contaminant of the cosmic microwave background (CMB) at frequencies above $\sim$100\,GHz, and its accurate separation is essential for current and future CMB experiments. Deep learning methods, in particular convolutional architectures such as the U-Net, have demonstrated strong performance for CMB component separation, but typically operate as purely data-driven models with no explicit enforcement of the physical laws governing the emission they reconstruct. In this work, we present a Physics-Informed Neural Network (PINN) that embeds the Modified Black Body (MBB) spectral model as a hard, differentiable architectural constraint, and we compare it against a standard U-Net baseline trained under an identical pipeline. Both architectures are trained on realistic simulated Planck HFI patches at four frequencies (143, 217, 353 and 545\,GHz) and evaluated on a shared validation set as well as on three independent Galactic dust realizations from the PySM phenomenological dust models named D4, D7 and D10. We find that both networks achieve comparable reconstruction quality across all evaluated datasets, with a Structural Similarity Index Measure consistently above 0.96 and a fitted angular power-spectrum spectral index that agrees with the input to within 1\% in every region. Embedding the physical constraint in the PINN does not measurably degrade reconstruction accuracy relative to the unconstrained U-Net, while additionally guaranteeing that every predicted output corresponds to a thermodynamically admissible physical configuration. These results establish the physics-informed formulation as a viable methodological alternative for dust component separation, providing physical interpretability without a trade-off in reconstruction fidelity.
\end{abstract}

\section{Introduction}

Artificial neural networks are universal function approximators, capable of learning highly non-linear mappings between high-dimensional input and output spaces directly from data, without an explicit closed-form model of the underlying process \citep{Goodfellow2016}. This property makes them particularly well suited to problems in which the forward physical model is well understood in principle but analytically intractable to invert at scale, as is the case for the separation of overlapping astrophysical emission components from multi-frequency observations.

The Interstellar Medium (ISM) is the diffuse matter that fills the space between stars within a galaxy, composed of gas, dust grains and the interstellar radiation field with which they interact. Interstellar dust grains, though a minor mass component of the ISM, dominate its opacity and emission in the infrared and submillimetre range, and their physical state (composition, size distribution and equilibrium temperature) is imprinted directly on the shape of their thermal emission spectrum.

In particular, interstellar dust is one of the most challenging ISM components to model quantitatively, since its emission arises from a population of grains embedded in a medium with strong non-linear, turbulent density and temperature fluctuations \citep{Draine2003, Draine2011}. This produces a spatially varying, non-Gaussian emission field whose parametrization, most commonly through the Modified Black Body (MBB) law, must be inferred locally rather than assumed constant across the sky, motivating a per-pixel estimation problem well suited to a convolutional, data-driven approach.

Furthermore, modeling correctly the interstellar dust emission is also relevant for Cosmic Microwave Background science, since it is one of the main contaminants for CMB recovery at frequencies above $\sim$100\,GHz \citep{PlanckXI2014}.

Interstellar dust grains absorb ultraviolet radiation from the interstellar radiation field and re-emit thermally in the far-infrared and submillimetre range \citep{Draine2003, Draine2011, Hensley2021}. Its spectral energy distribution is well described by the observations as a Modified Black Body (MBB) model:
\begin{equation}
    S(\nu) = A \cdot \nu^{\beta} \cdot B(\nu, T),
    \label{eq:mbb_intro}
\end{equation}
where $A$ is the emission amplitude, $\beta$ the spectral index and $T$ the equilibrium dust temperature, with $B(\nu,T)$ the Planck function. \textit{Planck} observations established global mean values of $\beta \approx 1.53$ and $T \approx 19.6$\,K, with significant spatial variations driven by local interstellar medium conditions \citep{PlanckXI2014, PlanckIntXXII2015}. A well-known anticorrelation between $\beta$ and $T$ complicates their simultaneous estimation, particularly at small angular scales \citep{Boudet2005, Meny2007, Paradis2011}.
Precise dust characterization is critical not only for CMB power spectrum recovery but especially for the detection of primordial B-mode polarization, whose amplitude may be orders of magnitude below the polarized dust signal \citep{BICEP2_2018, Tristram2020}. This requirement drives the design of next-generation experiments such as the Simons Observatory \citep{SimonsObservatory2019} and LiteBIRD \citep{Sugai2020}, which demand component separation methods of unprecedented accuracy and physical interpretability. Classical methods developed for \textit{Planck} are: NILC \citep{Delabrouille2009}, SEVEM \citep{MartinezGonzalez2003}, Commander \citep{Eriksen2006, Eriksen2008} and SMICA \citep{Delabrouille2003}, they all achieved excellent results but share inherent limitations. They require sky masking in the Galactic plane and near point sources, with the Planck confidence mask retaining approximately 78\% of the sky \citep{PlanckIV2020, Casas2022}. They also rely on parameterized foreground models and require instrument-specific calibration, making their adaptation to new experiments non-trivial.

Deep learning has emerged as a powerful alternative for component separation. One of the recent DL models was presented in Casas et al.\ \citep{Casas2022} developed CENN, a U-Net \citep{Ronneberger2015} trained on realistic \textit{Planck} simulations that recovers the CMB without applying any sky mask, outperforming classical methods in contaminated regions. This approach was subsequently extended to CMB polarization \citep{Casas2025} and to synchrotron separation in QUIJOTE data \citep{RubinoMartin2023}. Other methods such as Petroff et al.\ \citep{Petroff2020} and Yan et al.\ \citep{Yan2022} further demonstrated neural network viability for full-sky CMB recovery at multipoles up to $\ell \sim 1500$. For foreground modelling, generative approaches have been used to extend dust and synchrotron maps to sub-degree angular scales \citep{Krachmalnicoff2021, Thorne2021}.

All these methods, however, share a fundamental limitation: they learn a purely data-driven input--output mapping with no explicit enforcement of the physical emission laws governing the reconstructed signals, limiting both their interpretability and their ability to generalize beyond the training distribution.
Physics-Informed Neural Networks (PINNs), formalized by Raissi et al.\ \citep{Raissi2019}, address this by embedding physical knowledge directly into the network architecture or loss function, restricting solutions to physically admissible configurations and acting as a regularizer that amplifies the information content of the training data. Despite their success in fluid dynamics, materials science and geosciences \citep{Wang2023}, PINNs have not yet been applied to astrophysical component separation.

In this work we compare the performance of the baseline U-Net used in previous works with a new methodology based on PINNs with the aim of recovering a non-trivial and non-linear emission such as the Galactic thermal dust emission. In particular, the PINN incorporates the MBB model as a differentiable hard architectural constraint. Both networks reconstruct the dust emission at the same locations in the sky in the form of square patches of 25.6 degrees at 353 GHz.
We compare this architecture against a standard U-Net baseline, both trained on realistic Planck HFI simulations at four frequencies (143, 217, 353 and 545\,GHz) using the Planck Sky Model \citep{Delabrouille2013} and official FFP10 noise realizations. Generalization capability is assessed on three independent Galactic dust realizations combined with randomized instrumental noise, conditions unseen during training.

\section{Data}

The training and validation datasets consist of realistic simulations of the microwave sky at four Planck HFI frequencies: 143, 217, 353 and 545\,GHz. These frequencies were selected because the thermal dust contribution is significant across all of them, while the CMB dominates at lower frequencies. Each patch covers a $512 \times 512$ pixel field with a pixel size of 180\,arcsec. All simulations are projected onto flat patches using the zenithal equal-area (ZEA) projection implemented with \texttt{healpy} \citep{Zonca2019,Gorski2005},
\texttt{astropy} \citep{AstropyCollaboration2013,AstropyCollaboration2018,AstropyCollaboration2022}
and \texttt{reproject} \citep{Robitaille2023reproject}.

Each simulated sky patch at a given frequency is built as the sum of five components. The CMB contribution consists of intensity anisotropies from a $\Lambda$CDM realization with \textit{Planck} cosmological parameters, drawn from the Planck Legacy Archive (PLA). The Galactic thermal dust emission is simulated using dust maps from the Planck Sky Model (PSM) \citep{Delabrouille2013}, based on the GNILC reconstruction from the second Planck data release and extrapolated to each frequency assuming an MBB spectral law with spatially varying spectral index maps. Also, for each of the 10\,000 training simulations, a dust realization is selected randomly among three independent PySM dust models (D4, D7 and D10) \citep{Zonca2021,Thorne2017,Panexp2025}, each encoding different physical assumptions about the composition and spatial distribution of Galactic dust emission. D4 is a two-component MBB model, in which the total emission is the sum of two independent modified black bodies with different amplitudes, spectral indices and temperatures, representing a generalization of the single-component model to a mixture of cold and warm dust populations along the line of sight \citep{FinkbeinerDavisSchlegel1999}. D7 extends the single-component dust model of Hensley and Draine  \citep{HensleyDraine2017} by including iron inclusions in the grain composition, which modify the predicted frequency dependence of the emissivity. D10 is a single-component MBB model built from a GNILC needlet-based analysis of Planck data, offering reduced contamination from the Cosmic Infrared Background (CIB) and point sources relative to the Commander-based maps used in simpler dust models, with small-scale spectral index and dust temperature fluctuations added through the logpoltens formalism. This randomization exposes the network to a diverse range of dust morphologies and spectral behaviours during training, improving its generalization capability.

For the test datasets, the dust model is fixed: each of the three realizations defines an independent test set, split into high-latitude ($|b| > 30^\circ$) and low-latitude ($|b| < 30^\circ$) subsets of 1000 patches each, allowing a controlled evaluation of network performance on each individual dust model separately.

We have added point sources that are simulated as three populations generated with CORRSKY \citep{GonzalezNuevo2005}: radio galaxies dominant at 143 and 217\,GHz, late-type infrared galaxies, and proto-spheroidal galaxies forming the Cosmic Infrared Background (CIB), dominant at 353 and 545\,GHz. All the sky maps are convolved with the respective Planck HFI beam before co-addition. Instrumental noise is added by drawing, for each simulated patch, a noise realization at random from a set of 10 official Planck FFP10 noise maps available in the Planck Legacy Archive, one set per frequency channel. This randomization, applied consistently across training and test datasets, ensures sufficient noise diversity to prevent the network from overfitting to a specific noise pattern, while keeping the noise properties strictly within the realistic \textit{Planck} instrumental characteristics.

The training target (i.e. label) is the isolated thermal dust map at 353\,GHz, which is the standard pivot frequency for dust phenomenological models and the frequency at which dust dominates the Planck HFI sky. Beam convolution uses the official Planck HFI FWHM values: 7.22\,arcmin at 143\,GHz, 4.90\,arcmin at 217\,GHz, 4.92\,arcmin at 353\,GHz and 4.67\,arcmin at 545\,GHz \citep{PlanckI2020}. Diffuse components and noise are used directly from the PLA maps, which already incorporate beam convolution.

A total of 10\,000 sky patch simulations are used for training and 1000 independent patches for validation, with patch centre positions drawn randomly across the full celestial sphere for both sets. Generalization is assessed on six additional test datasets of 1000 patches each, corresponding to three independent Galactic dust realizations (D4, D7 and D10) split into high-latitude ($|b| > 30^\circ$) and low-latitude ($|b| < 30^\circ$) subsets, yielding a total of 17\,000 simulated patches across all datasets.

\section{Methodology}

\subsection{U-Net Architecture}

\begin{figure}[H]
\centering
\includegraphics[width=\linewidth]{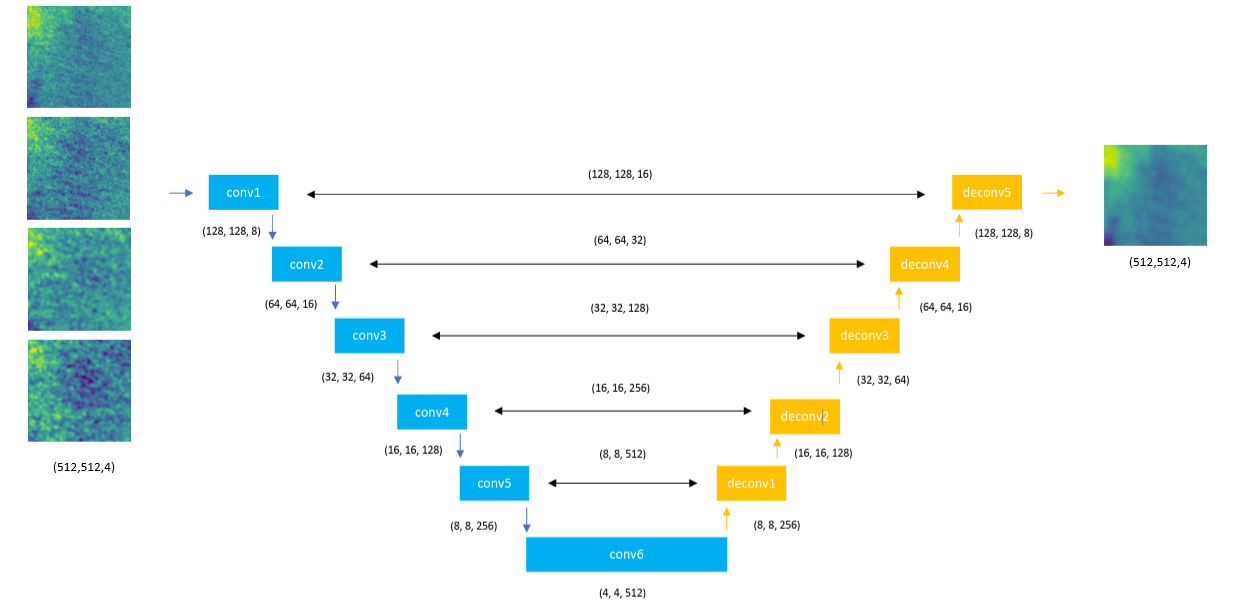}
\caption{Architecture of the U-Net baseline model. Blue blocks correspond to the encoder (convolutional blocks with downsampling) and orange blocks to the decoder (deconvolutional blocks with upsampling). Horizontal arrows represent skip connections that transmit fine-grained spatial information between symmetric encoder and decoder levels. The input consists of four frequency maps ($512\times512\times4$ channels) and the output is the reconstructed dust map at 353\,GHz ($512\times512\times1$ channel). Figure adapted from Casas et al.\ \citep{Casas2025}.}
\label{fig:unet_arch}
\end{figure}

The baseline model is a fully convolutional network based on the U-Net architecture \citep{Ronneberger2015}, following the design of the CMB Extraction Neural Network (CENN) developed by Casas et al.\ \citep{Casas2022, Casas2025}. The architecture consists of a symmetric encoder--decoder with six convolutional blocks in the encoder and six deconvolutional blocks in the decoder. Skip connections at each resolution level transmit fine-grained spatial information from the encoder to the decoder, allowing simultaneous recovery of structures at large and small angular scales. All activation functions are Leaky ReLU with slope $\alpha = 0.2$. The full architecture is illustrated in Figure~\ref{fig:unet_arch}.

The network takes as input the four frequency maps normalized to $[0,1]$ (tensor of shape $512 \times 512 \times 4$) and produces a single-channel output map (shape $512 \times 512 \times 1$) corresponding to the reconstructed thermal dust emission at 353\,GHz. The loss function is the Mean Squared Error (MSE) between the predicted and true dust maps:
\begin{equation}
    \mathcal{L}_{\rm U-Net} = \frac{1}{N} \sum_{i=1}^{N}
    \left( y_i - \hat{y}_i \right)^2,
    \label{eq:mse}
\end{equation}
where $N$ is the total number of pixels per patch, $y_i$ is the label value and $\hat{y}_i$ is the network prediction.

\begin{figure}[H]
\centering
\includegraphics[width=0.85\linewidth]{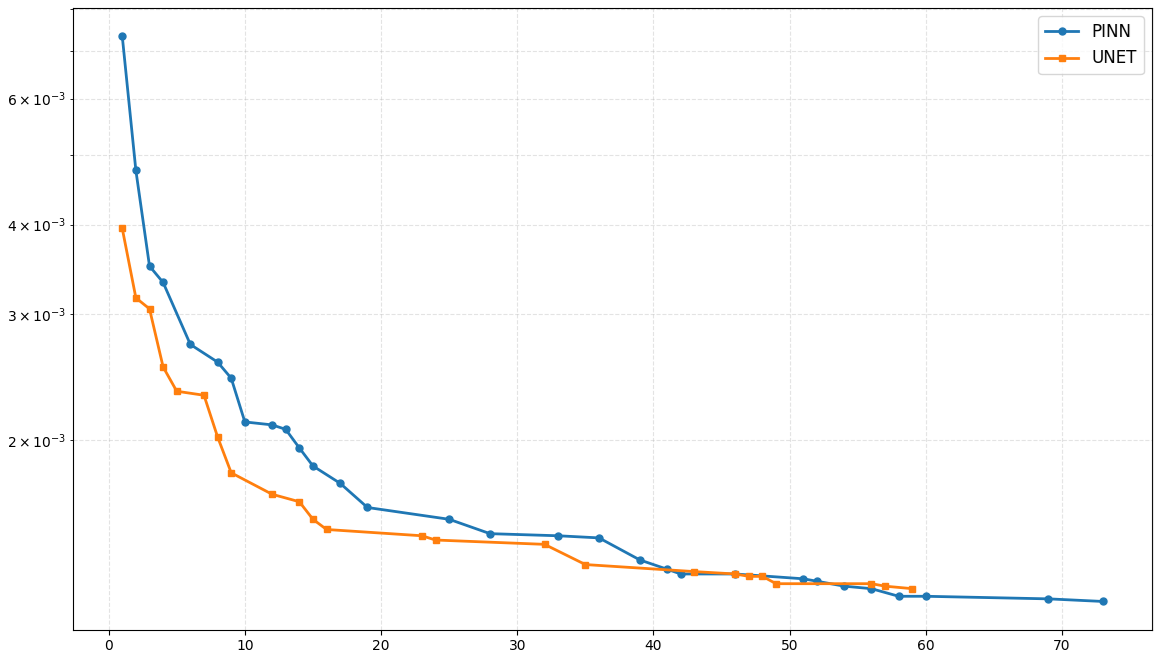}
\caption{Evolution of the validation loss (MSE) during training of the U-Net and the PINN, shown on a logarithmic vertical axis. The U-Net reaches its best validation loss of $1.24 \times 10^{-3}$ at epoch 59, a 68.6\% reduction relative to its initial loss of $3.95 \times 10^{-3}$. The PINN reaches its best validation loss of $1.19 \times 10^{-3}$ at epoch 73, an 83.8\% reduction relative to its initial loss of $7.33 \times 10^{-3}$. Both models converge to a validation MSE of comparable order of magnitude despite the PINN's additional physical constraint, and both show a higher absolute loss than in earlier training configurations, reflecting the increased difficulty of this dataset, in which the dust realization is randomized across three independent PySM models (D4, D7, D10) during training rather than fixed to a single model.}
\label{fig:loss_comparison}
\end{figure}

The network is trained with the Adam optimizer \citep{Kingma2015}, an initial learning rate of $10^{-4}$ and a mini-batch size of 8 patches, for a maximum of 300 epochs. The learning rate is halved whenever the validation loss shows no improvement for 10 consecutive epochs (down to a minimum of $10^{-6}$), and training is stopped early if the validation loss does not improve for 20 consecutive epochs, restoring the weights of the best-performing checkpoint. The model checkpoint corresponding to the lowest validation loss is retained as the final model. All training is performed with TensorFlow 2.x/Keras on an NVIDIA RTX 4070 GPU. Figure~\ref{fig:loss_comparison} shows the evolution of the validation loss during training for both the U-Net and the PINN.

Before quantitative evaluation, all predicted and label maps undergo two post-processing steps. First, a Gaussian smoothing kernel of 2.5\,arcmin FWHM is applied to mitigate pixelization artefacts introduced by the deconvolutional operations. Second, 10 pixels are cropped on each border to eliminate edge effects. Both steps are applied identically to predictions and labels.

\subsection{PINN Architecture}

\begin{figure}[H]
\centering
\includegraphics[width=\linewidth]{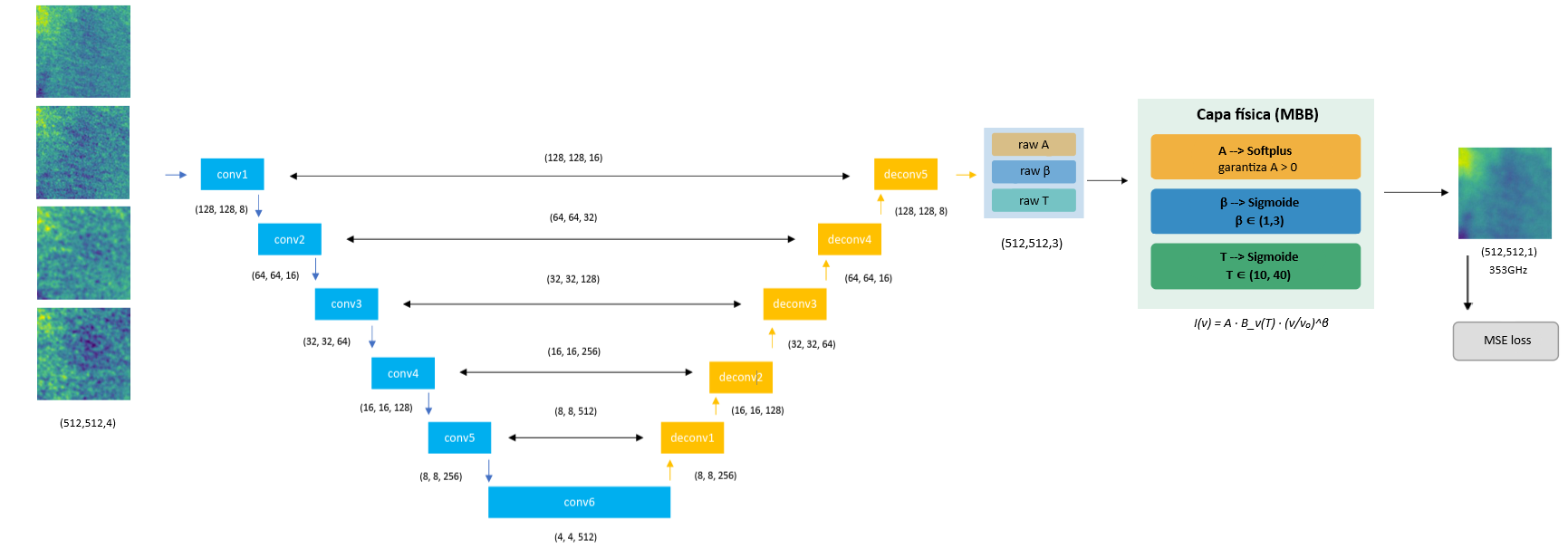}
\caption{Architecture of the proposed PINN. The left-hand side is identical to the U-Net baseline (Figure~\ref{fig:unet_arch}). The extension consists of replacing the single-channel output layer with three physical parameter maps (raw $A$, $\beta$, $T$), which are transformed through specialized activation functions (softplus for $A$, bounded sigmoid to $(1,3)$ for $\beta$, bounded sigmoid to $(10,40)$\,K for $T$) and combined in the non-trainable MBB layer to produce the reconstructed dust map at 353\,GHz.}
\label{fig:pinn_arch}
\end{figure}

The PINN extends the U-Net by replacing the single-channel output with a three-channel output that directly estimates the physical parameters of the Modified Black Body (MBB) spectral model: the emission amplitude $A$, the spectral index $\beta$ and the dust temperature $T$. A non-trainable MBB layer then combines these parameters to reconstruct the dust emission at 353\,GHz, embedding the physical emission law directly into the network graph. The full architecture is shown in Figure~\ref{fig:pinn_arch}.

The raw three-channel output $(\mathrm{raw}_A, \mathrm{raw}_\beta, \mathrm{raw}_T)$ is transformed through bounded activation functions to enforce thermodynamic constraints:
\begin{align}
    A      &= \mathrm{softplus}(\mathrm{raw}_A)
             = \log(1 + e^{\mathrm{raw}_A}),
             \label{eq:softplus}\\
    \beta  &= 1.0 + 2.0 \cdot \sigma(\mathrm{raw}_\beta),
             \label{eq:beta}\\
    T      &= 10.0 + 30.0 \cdot \sigma(\mathrm{raw}_T),
             \label{eq:temp}
\end{align}
where $\sigma(\cdot)$ denotes the standard sigmoid function. These transformations guarantee $A > 0$, $\beta \in (1, 3)$ and $T \in (10, 40)$\,K, ranges that encompass the full range of physically observed dust properties \citep{PlanckXI2014, PlanckIntXXII2015}.

The dust emission map at 353\,GHz is then reconstructed via the MBB equation:
\begin{equation}
    S(\nu_{353}) = A \cdot \nu_{353}^{\beta} \cdot B(\nu_{353}, T),
    \label{eq:mbb}
\end{equation}
where $B(\nu, T) = \frac{2h\nu^3}{c^2}\frac{1}{e^{h\nu/k_BT}-1}$ is the Planck function evaluated at $\nu_{353} = 353 \times 10^9$\,Hz, $h$ is the Planck constant, $k_B$ the Boltzmann constant and $c$ the speed of light.

The MBB layer is fully differentiable, enabling gradient propagation from the MSE loss through the physical model to the encoder--decoder weights. The loss function is identical in form to that of the U-Net (Equation~\ref{eq:mse}), with $\hat{y}_i$ replaced by $S_i(\nu_{353})$. This implementation constitutes a \emph{hard-constraint} PINN formulation: physical consistency is enforced by architecture rather than by a penalty term in the loss, ensuring that every prediction lies within the thermodynamically admissible parameter space without requiring manual weighting of competing loss terms.

The PINN is trained under the same procedure as the U-Net: the Adam optimizer \citep{Kingma2015} with an initial learning rate of $10^{-4}$, a mini-batch size of 8 patches and a maximum of 300 epochs, with the learning rate halved after 10 epochs without improvement in the validation loss (down to a minimum of $10^{-6}$) and early stopping after 20 epochs without improvement, restoring the best-performing weights. The same post-processing pipeline described above (2.5\,arcmin Gaussian smoothing and 10-pixel border cropping) is applied identically to the PINN predictions and labels prior to evaluation.

Reconstruction quality is assessed with four complementary spatial-domain metrics, Root Mean Squared Error (RMSE), Mean Absolute Error (MAE), Structural Similarity Index Measure (SSIM) and Peak Signal-to-Noise Ratio (PSNR), computed on denormalized maps after post-processing and reported as mean $\pm$ standard deviation over all patches in each test set. RMSE and MAE quantify pixel-wise reconstruction error in physical units and are directly comparable between the two networks, since both are trained to minimize a loss of the same form (Equation~\ref{eq:mse}). SSIM assesses whether the large-scale morphological structure of the dust emission, such as filamentary features and extended diffuse regions, is preserved independently of absolute amplitude differences, while PSNR provides a complementary dynamic-range-normalized measure of reconstruction quality, particularly informative given the several-orders-of-magnitude variation in dust intensity between low- and high-Galactic-latitude regions. In addition, the angular power spectrum $C_\ell$ is estimated for each patch via 2D Fast Fourier Transform, with mean subtraction and a 2D Hann window to suppress spectral leakage. Multipoles are computed as $\ell = 2\pi|k|$ and binned with adaptive bin widths over the range $300 \leq \ell \leq 2000$. The mean spectrum is computed over all patches in each test set, with uncertainty estimated as the standard error of the mean $\sigma/\sqrt{N}$, and a power law $C_\ell = A_d \cdot \ell^{-\alpha_d}$ is fitted to the mean spectrum for both input and output in each sky region, with the relative difference in spectral index $\Delta\alpha_d$ used as an independent measure of spectral fidelity. Together, these diagnostics assess reconstruction fidelity at the level of pixel error, structural morphology, dynamic range and spectral content, testing whether the networks preserve the distribution of power across angular scales in addition to pixel-domain amplitude, a property not captured by the spatial-domain metrics alone.

\section{Results}

\subsection{Validation with baseline D10 model}
\label{sec:validation_simulations}

Prior to network input, all patches are normalized independently per channel using a min-max normalization:
\begin{equation}
    x_{\rm norm} = \frac{x - x_{\min}}{x_{\max} - x_{\min}},
    \label{eq:minmax}
\end{equation}
where $x_{\min}$ and $x_{\max}$ denote the minimum and maximum pixel values within each channel and patch. The normalization parameters are stored alongside each simulation to allow denormalization of network predictions before computing the metrics reported below.

\begin{table}[H]
\centering
\caption{Reconstruction metrics (mean $\pm$ standard deviation) for the PINN and U-Net across the validation set and all generalization test datasets. The root mean squared error (RMSE) and mean absolute error (MAE)  quantify the pixel-wise reconstruction error and are given in units of $K_{\rm CMB}$, with lower values indicating better agreement with the target map. The structural similarity index measure (SSIM) quantifies the preservation of spatial structures, with values closer to 1 indicating greater structural similarity. The peak signal-to-noise ratio (PSNR) measures the reconstruction quality on a logarithmic scale and is given in dB, with higher values indicating better reconstruction quality. Each dataset consists of 1000 patches. The validation set shares the dust model distribution of the training set, while each generalization dataset is evaluated on a single fixed dust realization (D4, D7 or D10) split by Galactic latitude.}
\label{tab:reconstruction}
\resizebox{\linewidth}{!}{%
\begin{tabular}{llcccc}
\hline
\textbf{Model} & \textbf{Dataset} & \textbf{RMSE} & \textbf{MAE}
& \textbf{SSIM} & \textbf{PSNR [dB]} \\
\hline
\multirow{7}{*}{PINN}
 & Validation   & $2.01\times10^{-4} \pm 5.26\times10^{-4}$
                & $1.60\times10^{-4} \pm 4.42\times10^{-4}$
                & $0.974 \pm 0.033$ & $35.99 \pm 7.82$ \\
 & D4 high lat  & $4.01\times10^{-5} \pm 9.23\times10^{-5}$
                & $3.33\times10^{-5} \pm 8.95\times10^{-5}$
                & $0.967 \pm 0.039$ & $33.29 \pm 6.62$ \\
 & D4 low lat   & $5.33\times10^{-4} \pm 8.14\times10^{-4}$
                & $4.31\times10^{-4} \pm 6.86\times10^{-4}$
                & $0.985 \pm 0.027$ & $42.43 \pm 4.08$ \\
 & D7 high lat  & $3.59\times10^{-5} \pm 3.23\times10^{-5}$
                & $2.94\times10^{-5} \pm 2.85\times10^{-5}$
                & $0.972 \pm 0.027$ & $34.77 \pm 5.86$ \\
 & D7 low lat   & $5.43\times10^{-4} \pm 8.89\times10^{-4}$
                & $4.45\times10^{-4} \pm 7.69\times10^{-4}$
                & $0.985 \pm 0.026$ & $42.68 \pm 3.63$ \\
 & D10 high lat & $4.07\times10^{-5} \pm 2.16\times10^{-5}$
                & $3.32\times10^{-5} \pm 1.87\times10^{-5}$
                & $0.966 \pm 0.030$ & $29.37 \pm 6.76$ \\
 & D10 low lat  & $3.01\times10^{-4} \pm 3.37\times10^{-4}$
                & $2.17\times10^{-4} \pm 2.46\times10^{-4}$
                & $0.995 \pm 0.004$ & $41.92 \pm 5.61$ \\
\hline
\multirow{7}{*}{U-Net}
 & Validation   & $1.75\times10^{-4} \pm 4.30\times10^{-4}$
                & $1.48\times10^{-4} \pm 3.84\times10^{-4}$
                & $0.970 \pm 0.038$ & $36.21 \pm 8.25$ \\
 & D4 high lat  & $3.80\times10^{-5} \pm 7.08\times10^{-5}$
                & $3.11\times10^{-5} \pm 6.84\times10^{-5}$
                & $0.965 \pm 0.056$ & $33.25 \pm 6.92$ \\
 & D4 low lat   & $4.38\times10^{-4} \pm 6.37\times10^{-4}$
                & $3.73\times10^{-4} \pm 5.73\times10^{-4}$
                & $0.975 \pm 0.054$ & $43.64 \pm 4.88$ \\
 & D7 high lat  & $3.42\times10^{-5} \pm 2.36\times10^{-5}$
                & $2.75\times10^{-5} \pm 1.97\times10^{-5}$
                & $0.971 \pm 0.028$ & $34.67 \pm 6.23$ \\
 & D7 low lat   & $4.36\times10^{-4} \pm 7.04\times10^{-4}$
                & $3.71\times10^{-4} \pm 6.37\times10^{-4}$
                & $0.977 \pm 0.044$ & $44.19 \pm 4.29$ \\
 & D10 high lat & $4.25\times10^{-5} \pm 2.35\times10^{-5}$
                & $3.46\times10^{-5} \pm 1.99\times10^{-5}$
                & $0.965 \pm 0.030$ & $29.03 \pm 6.68$ \\
 & D10 low lat  & $3.00\times10^{-4} \pm 2.94\times10^{-4}$
                & $2.34\times10^{-4} \pm 2.39\times10^{-4}$
                & $0.992 \pm 0.008$ & $41.59 \pm 6.09$ \\
\hline
\end{tabular}
}
\end{table}

Table~\ref{tab:reconstruction} presents the full quantitative reconstruction metrics for both architectures across all evaluated datasets. On the validation set, the U-Net reaches SSIM $= 0.970 \pm 0.038$ and PSNR $= 36.2 \pm 8.2$\,dB, while the PINN achieves SSIM $= 0.974 \pm 0.033$ and PSNR $= 36.0 \pm 7.8$\,dB. The U-Net presents marginally lower RMSE ($1.75 \times 10^{-4}$ vs.\ $2.01 \times 10^{-4}$), consistent with the additional constraint imposed by the MBB layer, which restricts the PINN solution space to physically admissible configurations. The difference in SSIM between both architectures is below 0.005 on the validation set, confirming that the physical constraint does not penalize morphological reconstruction quality.

\begin{figure}[H]
\centering
\begin{minipage}{\linewidth}
\centering
\includegraphics[width=\linewidth]{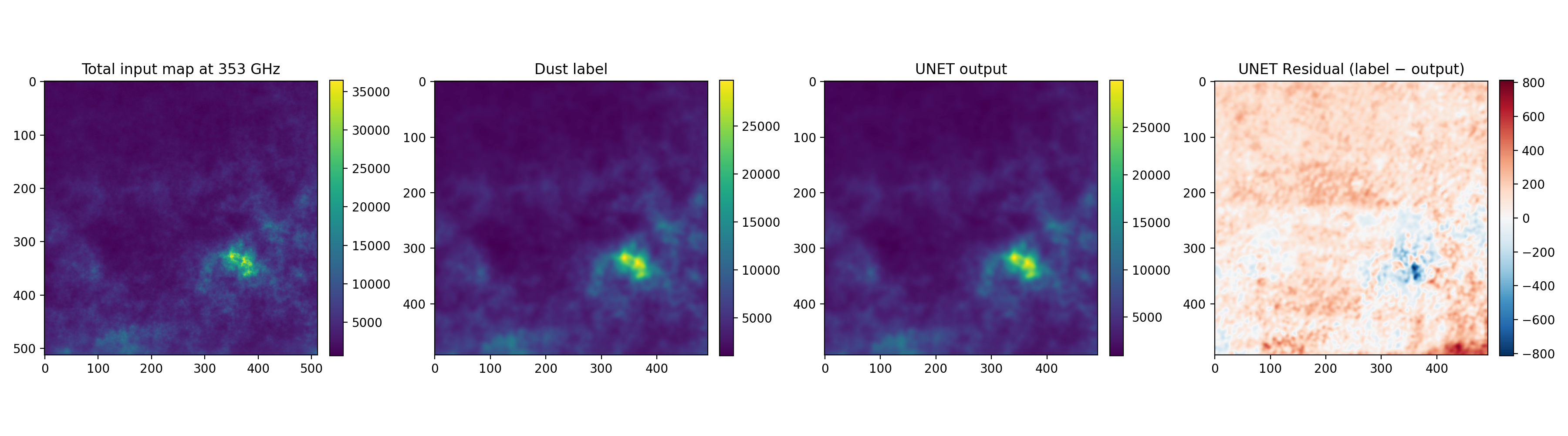}
\end{minipage}
\begin{minipage}{\linewidth}
\centering
\includegraphics[width=\linewidth]{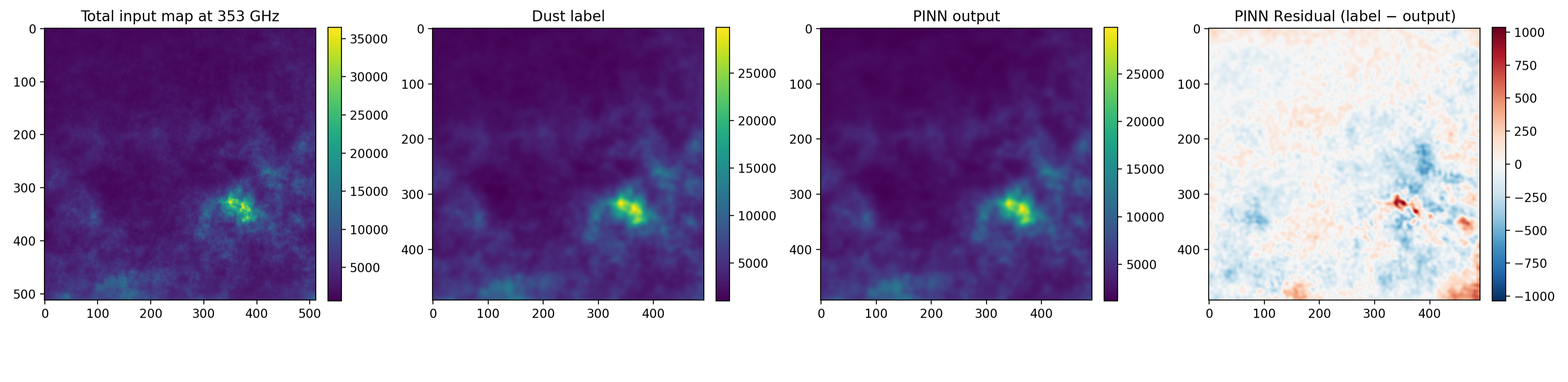}
\end{minipage}
\caption{An example of the reconstruction of a low-latitude patch ($|b| < 30^\circ$) from a sky with the D10 dust emission by the U-Net (top) and the PINN (bottom). From left to right: total input map with all the emissions at 353\,GHz, isolated dust emission, used as a label for the training, the output from each network and the residuals. Colorbars show the $\mu K_{\mathrm{CMB}}$ units for each patch.}
\label{fig:cl_d10_low_patch}
\end{figure}

\begin{figure}[H]
\centering
\begin{minipage}{\linewidth}
\centering
\includegraphics[width=\linewidth]{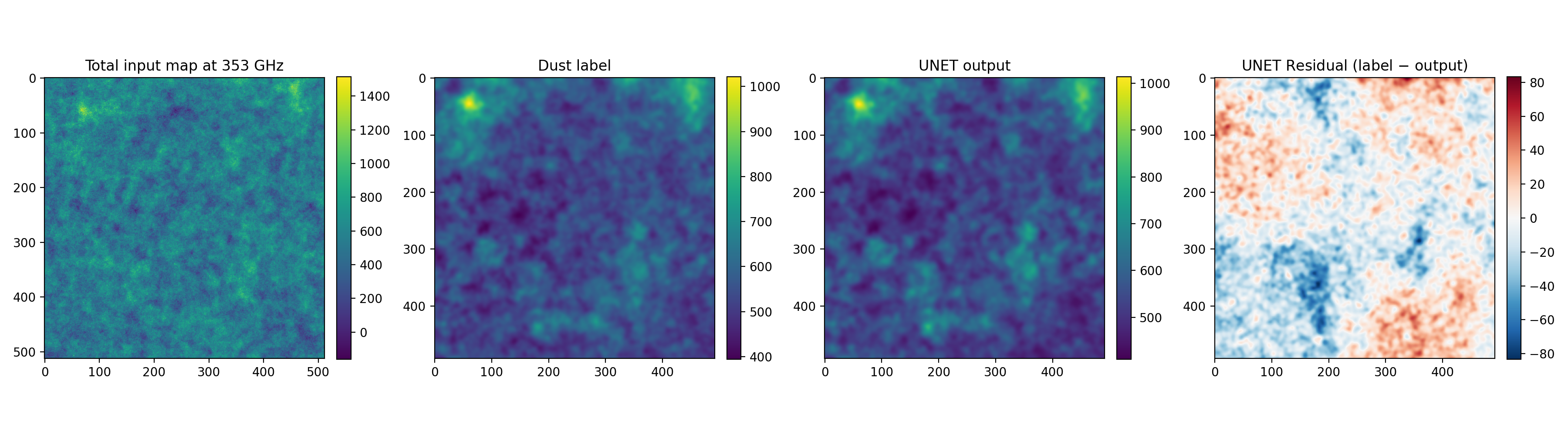}
\end{minipage}
\begin{minipage}{\linewidth}
\centering
\includegraphics[width=\linewidth]{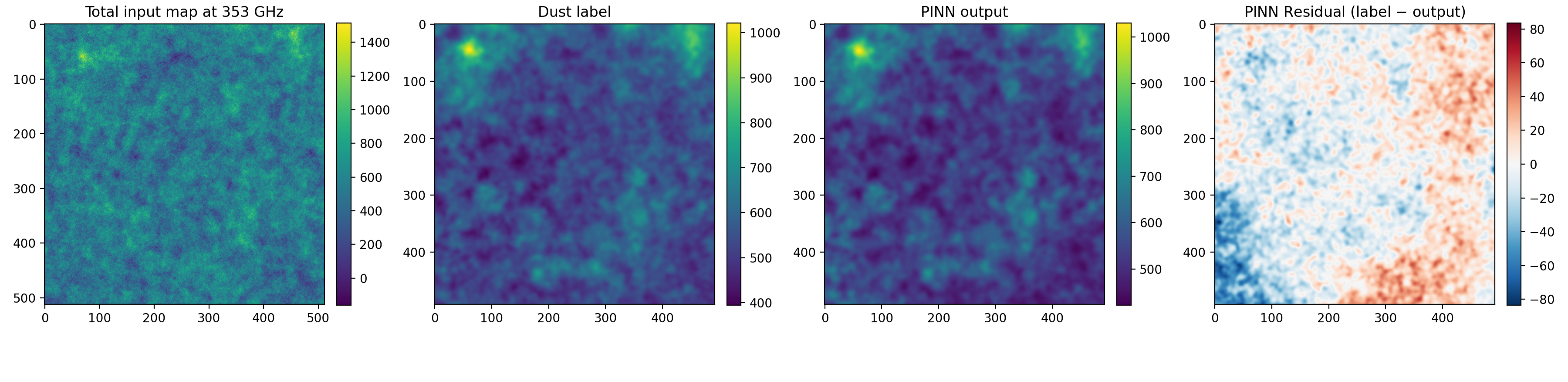}
\end{minipage}
\caption{An example of the reconstruction of a high-latitude patch ($|b| > 30^\circ$) from a sky with the D10 dust emission. Rows and columns represent the same as in Figure \ref{fig:cl_d10_low_patch}.}
\label{fig:cl_d10_high_patch}
\end{figure}

Representative patch reconstructions for low-latitude and high-latitude regions are shown in Figures~\ref{fig:cl_d10_low_patch} and~\ref{fig:cl_d10_high_patch}. Both figures compare the U-Net (top) and PINN (bottom) reconstructions of the same representative patches, showing from left to right the total input map at 353\,GHz, the isolated dust label, the network output, and the residual map. In the low-latitude patch (Figure~\ref{fig:cl_d10_low_patch}), a compact bright source is embedded in an extended diffuse background, the U-Net residual is dominated by low-amplitude speckle with a localized negative feature adjacent to the compact source, while the PINN residual spans a wider range and shows an additional coherent negative band running through the patch together with a concentrated residual at the location of the bright source itself. In the high-latitude patch (Figure~\ref{fig:cl_d10_high_patch}), where a single bright point source appears in one corner against a much fainter diffuse background, both networks show a large-scale coherent residual pattern across the patch, with negative and positive regions occupying opposite halves of the field, rather than residuals confined only to the vicinity of the point source.

\begin{figure}[H]
\centering
\begin{minipage}{0.49\linewidth}
\centering
\includegraphics[width=\linewidth]{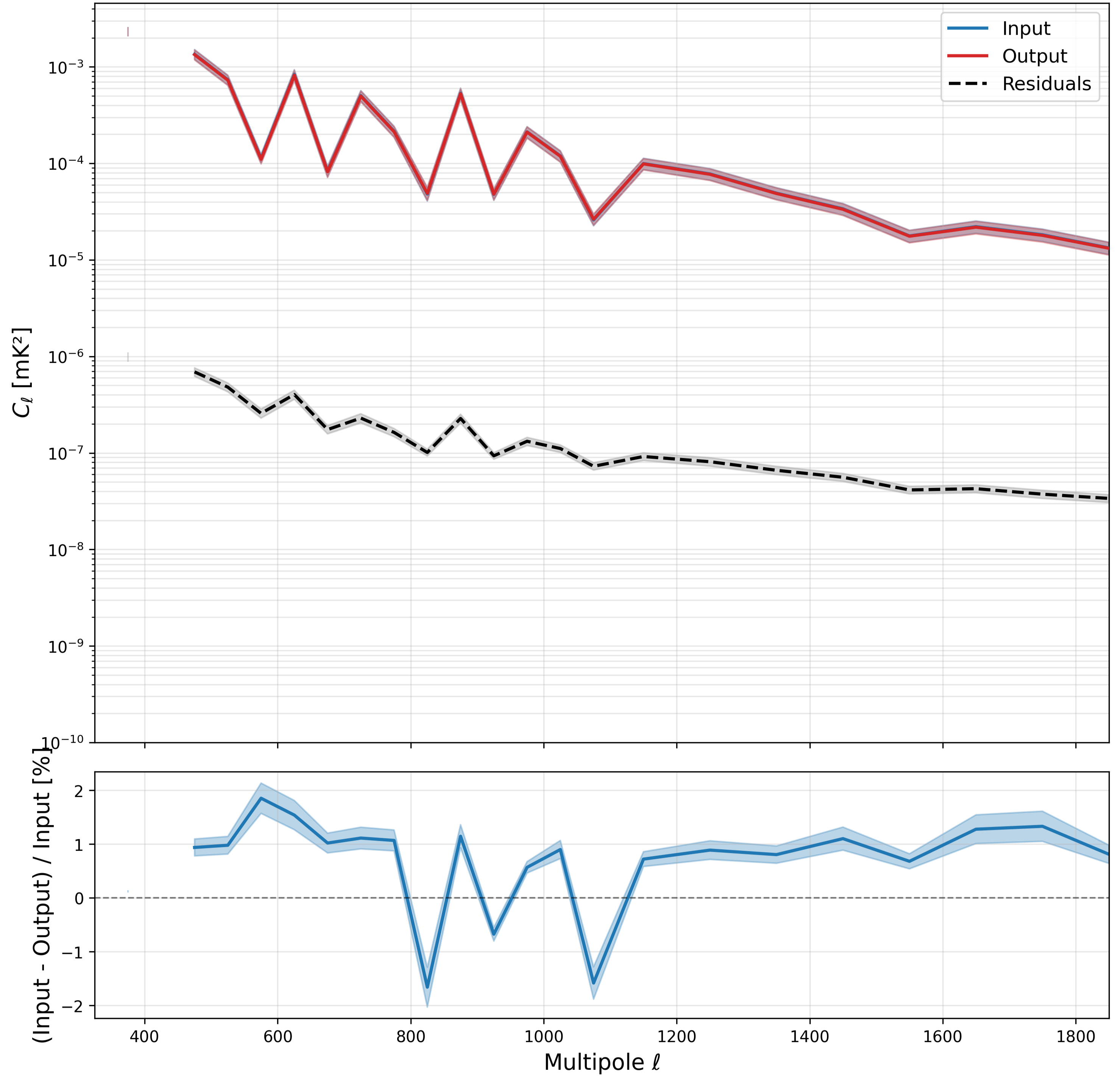}
\end{minipage}
\hfill
\begin{minipage}{0.49\linewidth}
\centering
\includegraphics[width=\linewidth]{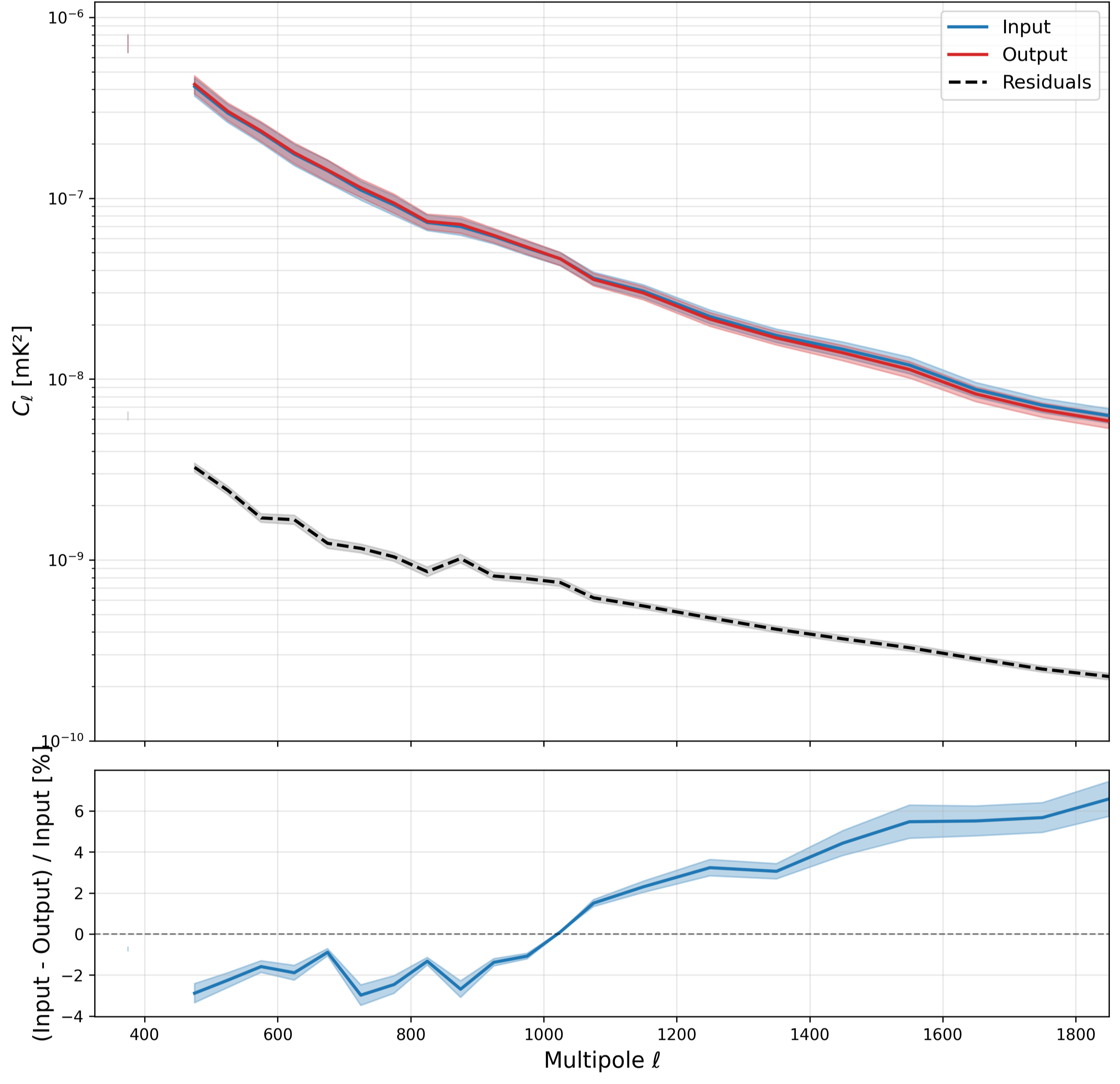}
\end{minipage}
\caption{Angular power spectrum $C_\ell$ of the dust emission reconstructed by the PINN for the D10 dust realization, comparing the Galactic-plane region (left panel, $|b| < 30^\circ$) and the region outside the Galactic plane (right panel, $|b| > 30^\circ$). In both panels, the upper plot shows the mean power spectra of the input label (blue), network output (red) and the residuals (dashed black), with $1\sigma$ uncertainty bands, while the lower plot shows the relative error. The comparison illustrates the degradation of reconstruction fidelity as the dust signal weakens away from the Galactic plane.}
\label{fig:pinn_cl_d10_lowhigh}
\end{figure}

\begin{figure}[H]
\centering
\begin{minipage}{0.49\linewidth}
\centering
\includegraphics[width=\linewidth]{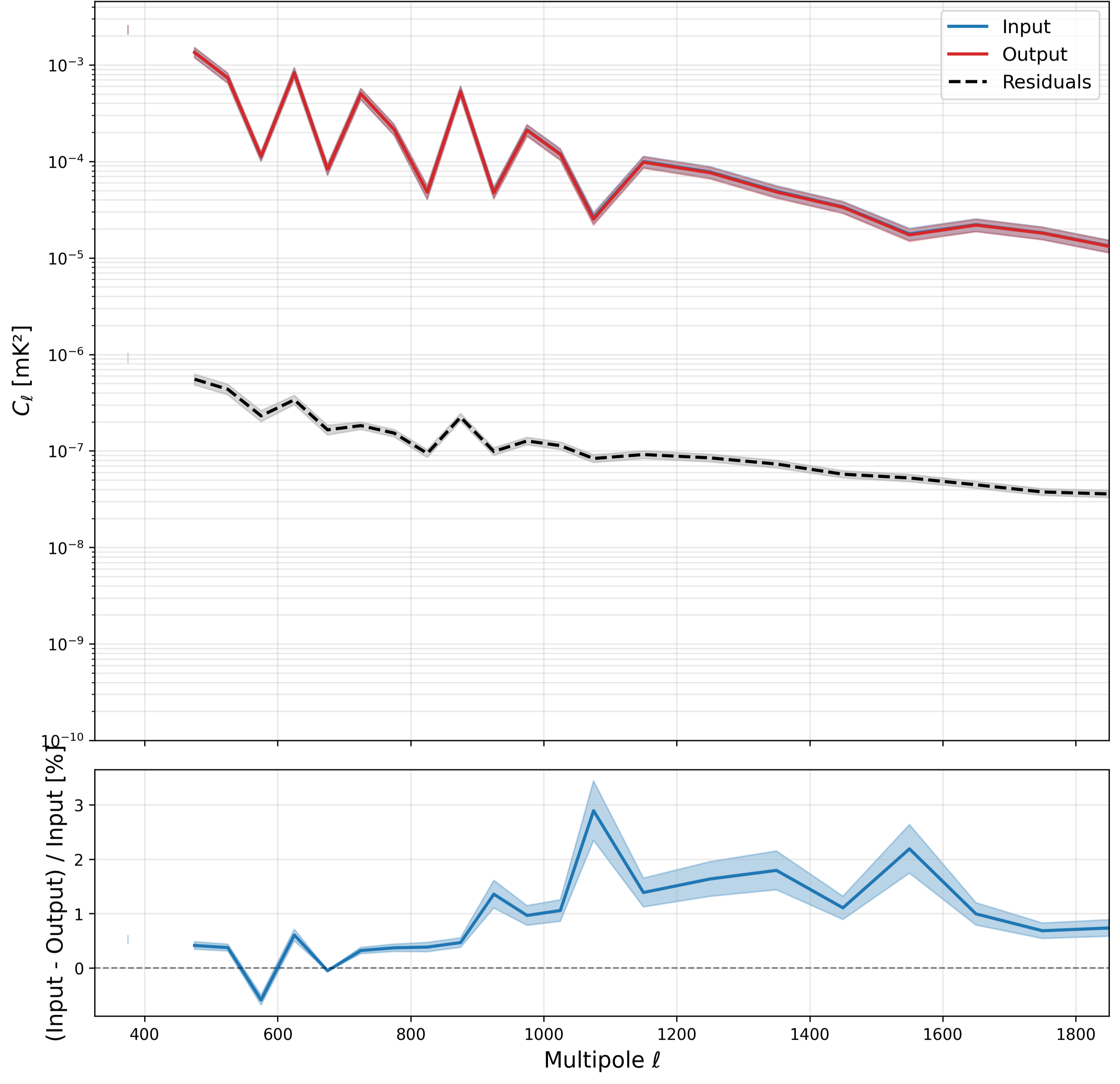}
\end{minipage}
\hfill
\begin{minipage}{0.49\linewidth}
\centering
\includegraphics[width=\linewidth]{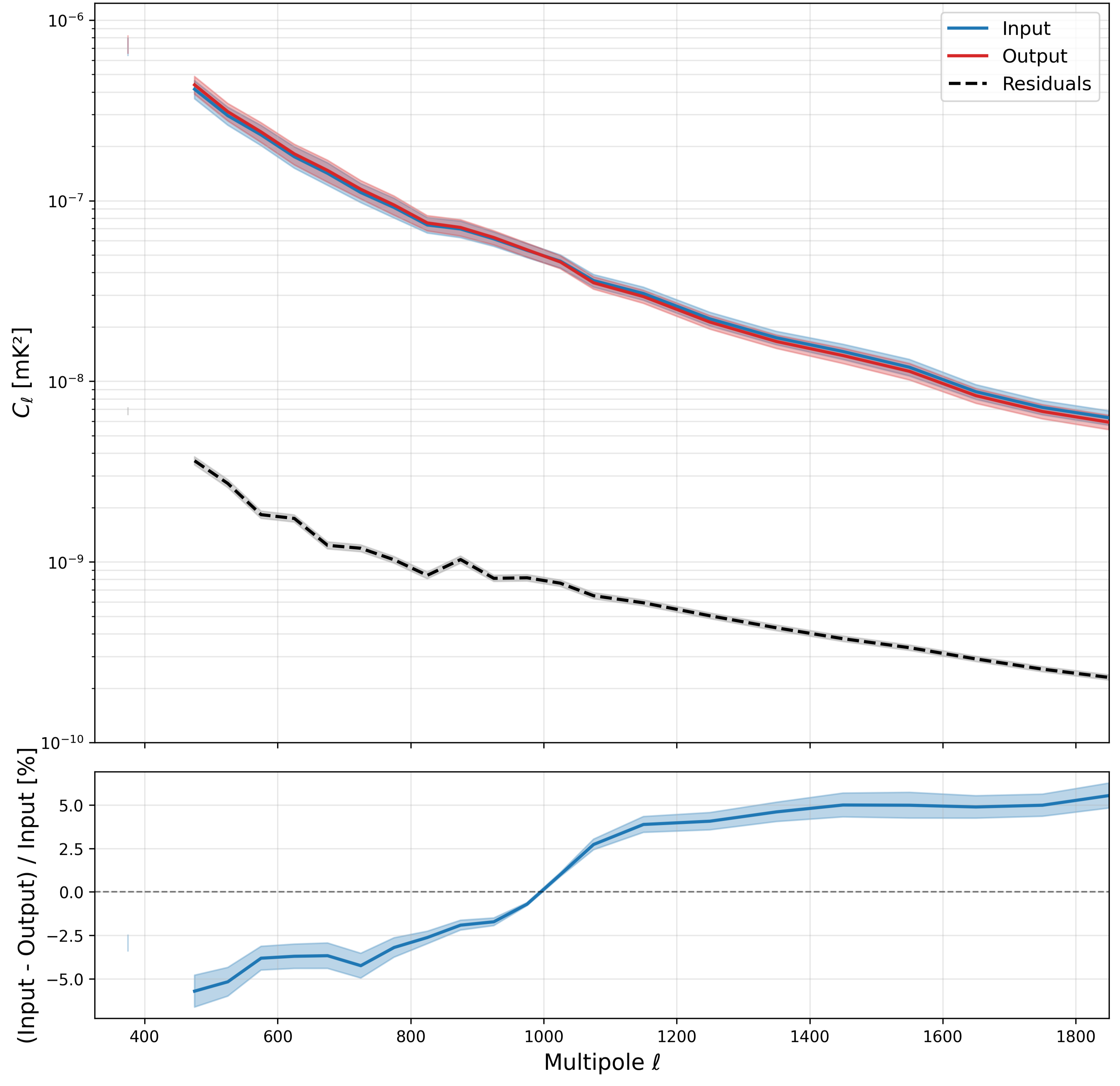}
\end{minipage}
\caption{Angular power spectrum $C_\ell$ of the dust emission reconstructed by the U-Net for the D10 dust realization, comparing the Galactic-plane region (left panel, $|b| < 30^\circ$) and the region outside the Galactic plane (right panel, $|b| > 30^\circ$). Panel layout identical to Figure~\ref{fig:pinn_cl_d10_lowhigh}.}
\label{fig:unet_cl_d10_lowhigh}
\end{figure}

The angular power spectrum analysis for the D10 dust realization is presented in Figures~\ref{fig:pinn_cl_d10_lowhigh} and~\ref{fig:unet_cl_d10_lowhigh}, comparing the PINN and U-Net respectively across the low- and high-latitude regions. Both networks reproduce the dust power spectrum with high fidelity across the multipole range $300 \leq \ell \leq 2000$. At low Galactic latitude, the uncertainty bands on the mean spectrum are visibly wider at the largest angular scales ($\ell \lesssim 500$) than at higher multipoles, reflecting greater patch-to-patch variability in this regime, this behavior is consistent between the input label and both network outputs, indicating that it originates from the underlying variability of the dust signal itself rather than from the reconstruction. The relative error between input and output power spectra remains within $\pm 3\%$ for the PINN and within $\pm 4\%$ for the U-Net across both latitude regions.

\begin{figure}[H]
\centering
\begin{minipage}{0.49\linewidth}
    \centering
    \includegraphics[width=\linewidth]{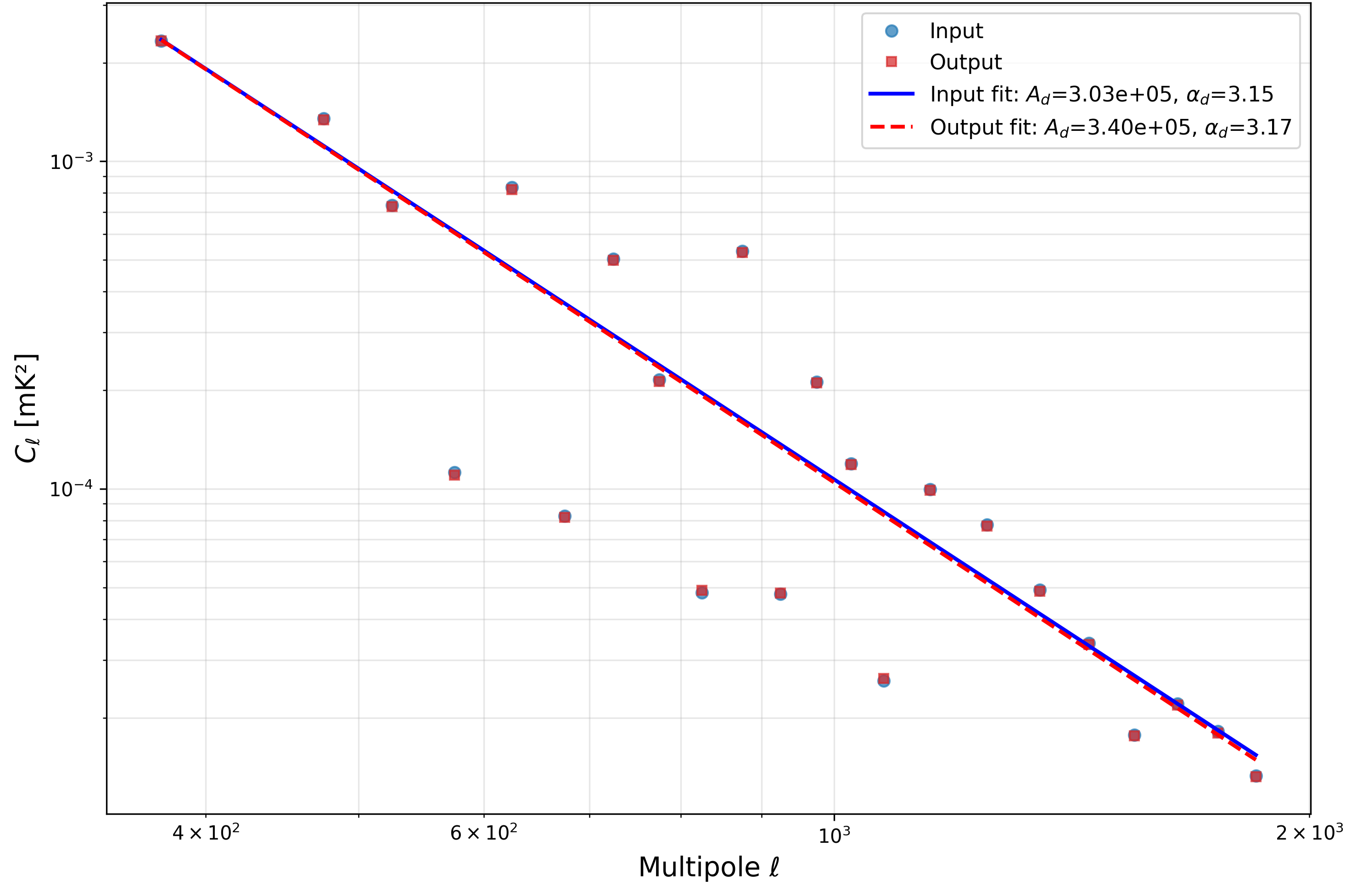}
\end{minipage}
\hfill
\begin{minipage}{0.49\linewidth}
    \centering
    \includegraphics[width=\linewidth]{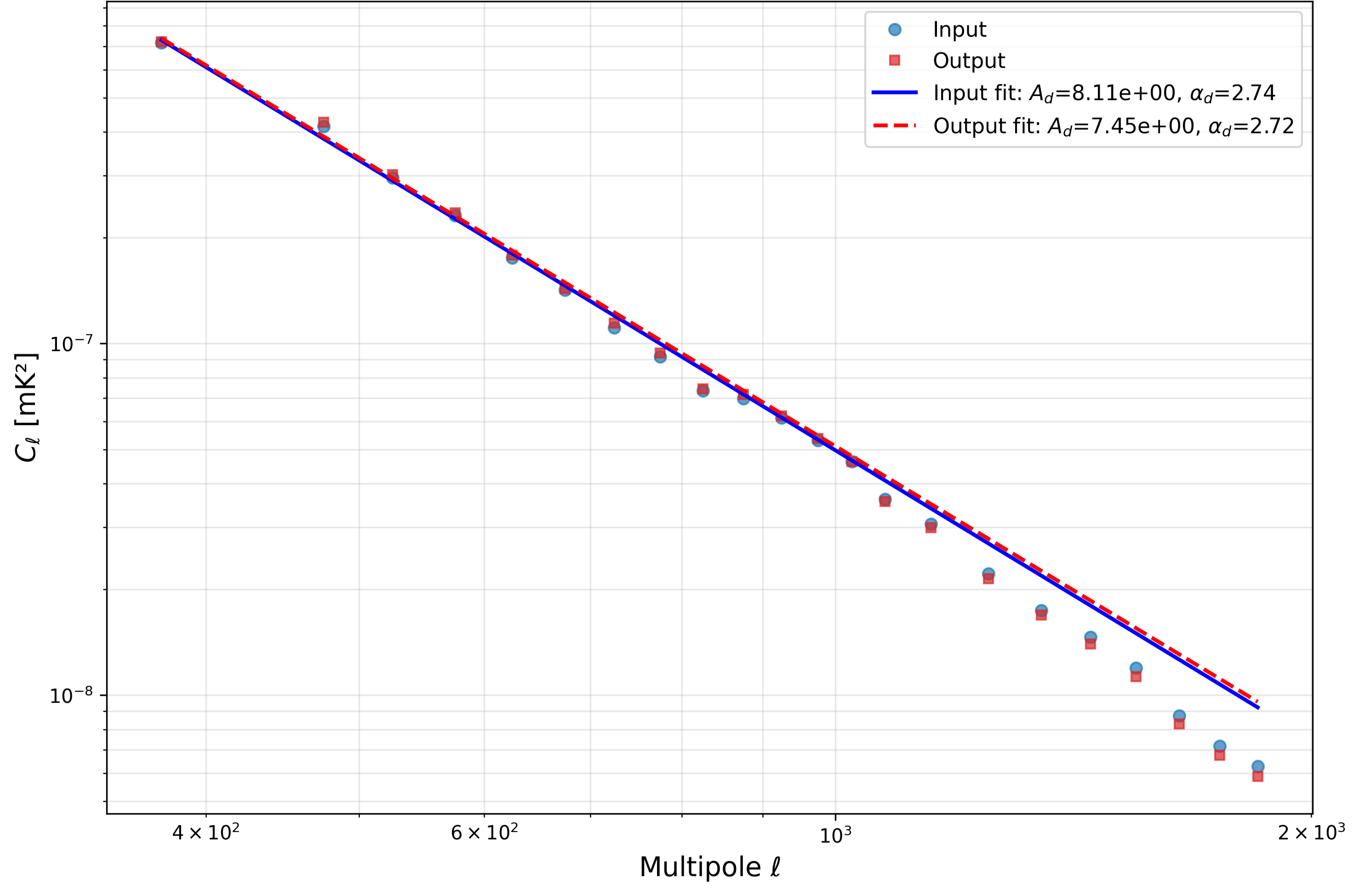}
\end{minipage}
\caption{Power-law fit $C_\ell = A_d\,\ell^{-\alpha_d}$ to the mean dust power spectrum for the PINN, D10 realization, comparing the low-latitude region (left panel, $|b| < 30^\circ$) and the high-latitude region (right panel, $|b| > 30^\circ$). Blue points and line: input label spectrum and fit. Red points and line: PINN output spectrum and fit. The fitted spectral indices agree to within $0.637\%$ (left panel) and $0.591\%$ (right panel).}
\label{fig:pinn_pl_d10_lowhigh}
\end{figure}

\begin{figure}[H]
\centering
\begin{minipage}{0.49\linewidth}
\centering
\includegraphics[width=\linewidth]{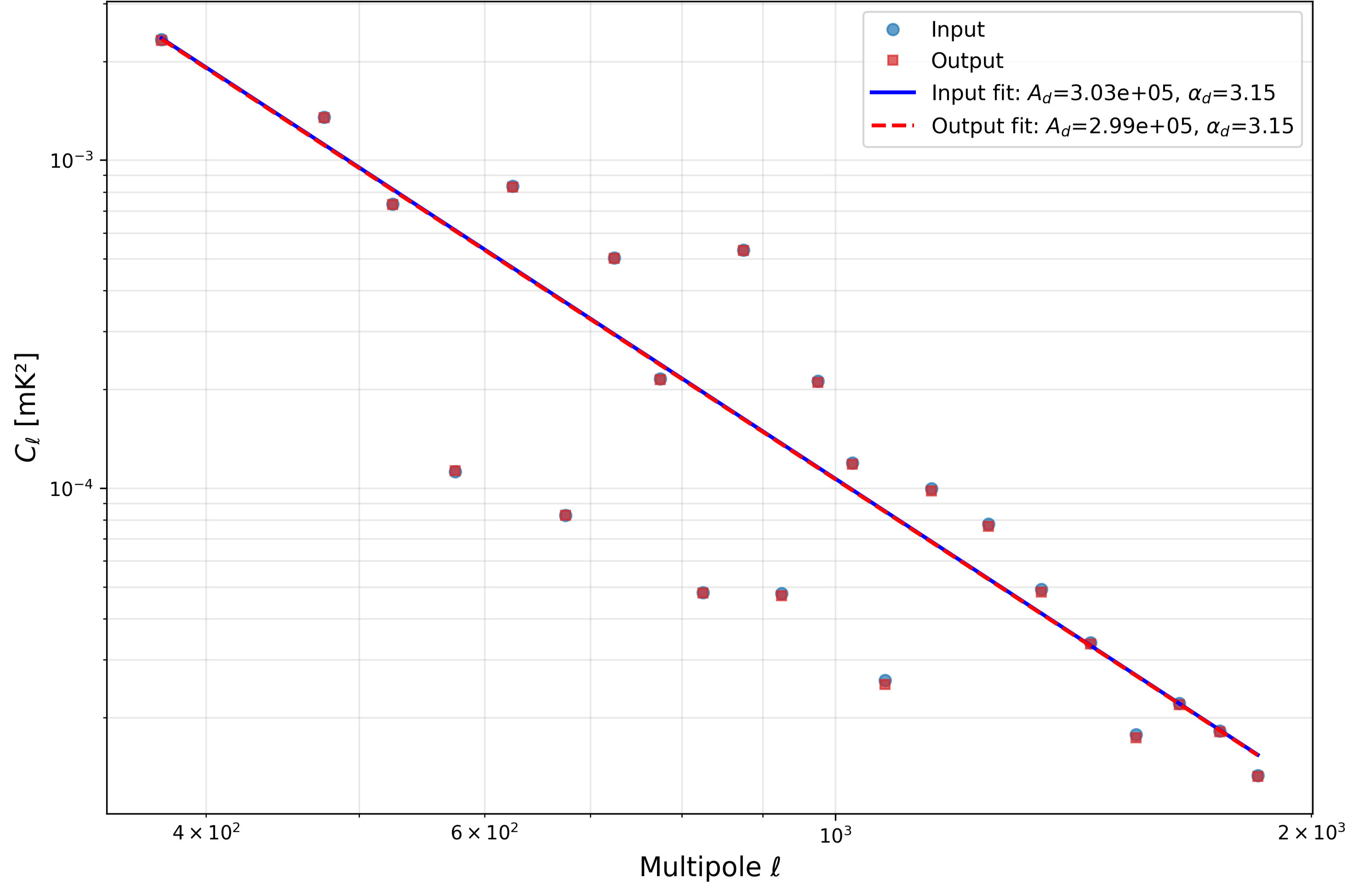}
\end{minipage}
\hfill
\begin{minipage}{0.49\linewidth}
\centering
\includegraphics[width=\linewidth]{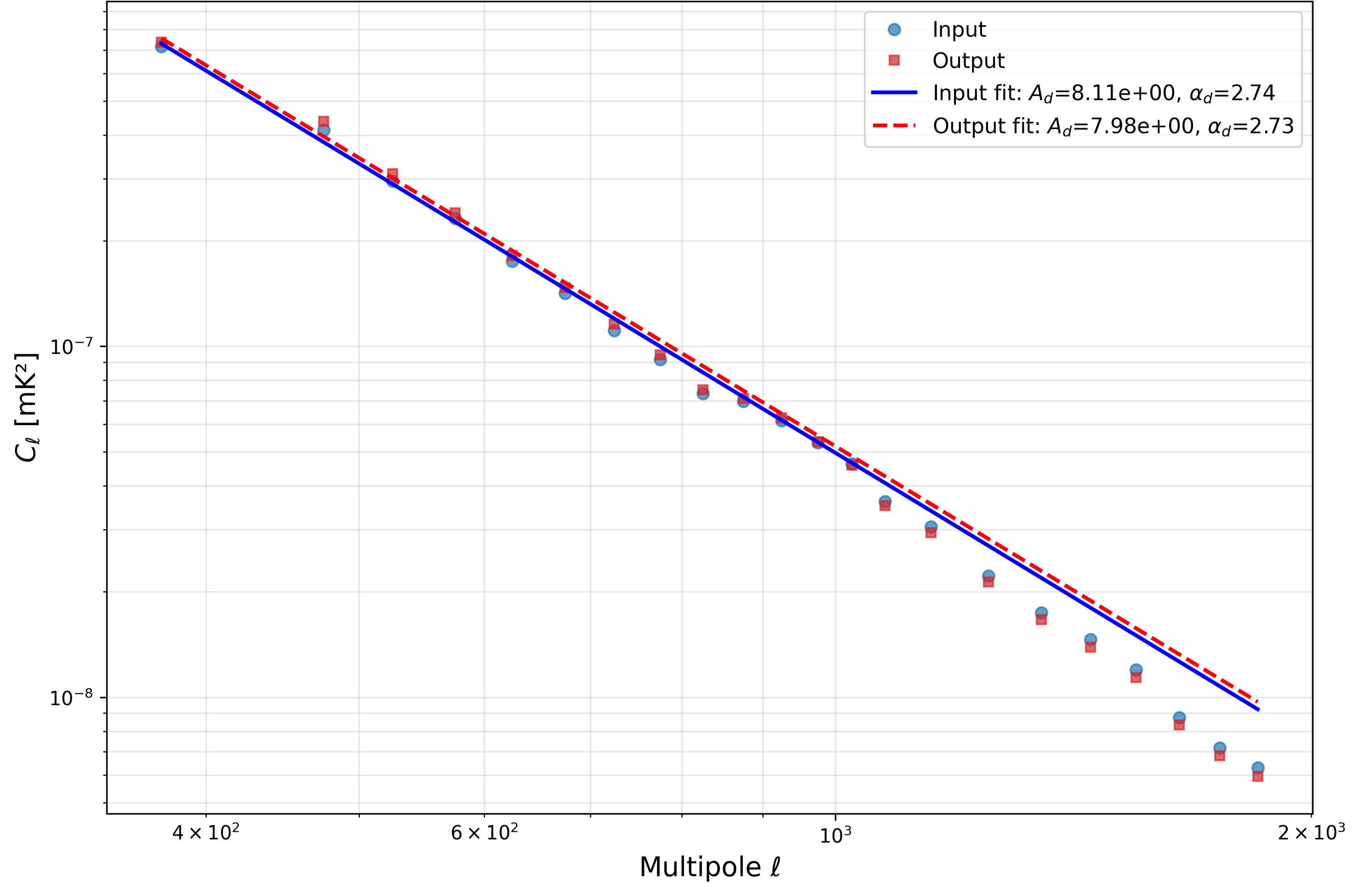}
\end{minipage}
\caption{Power-law fit $C_\ell = A_d\,\ell^{-\alpha_d}$ to the mean dust power spectrum for the U-Net, D10 realization, comparing the low-latitude region (left panel, $|b| < 30^\circ$) and the high-latitude region (right panel, $|b| > 30^\circ$). Blue points and line: input label spectrum and fit. Red points and line: U-Net output spectrum and fit. The fitted spectral indices agree to within $0.035\%$ (left panel) and $0.315\%$ (right panel).}
\label{fig:unet_pl_d10_lowhigh}
\end{figure}

The corresponding power-law fits are shown in Figures~\ref{fig:pinn_pl_d10_lowhigh} and~\ref{fig:unet_pl_d10_lowhigh}. The fitted power-law spectral index differs between input and output by $\Delta\alpha_d = 0.591\%$ and $\Delta\alpha_d = 0.637\%$ for the PINN at high and low latitudes respectively, and $\Delta\alpha_d = 0.315\%$ and $\Delta\alpha_d = 0.035\%$ for the U-Net, all well below 1\% (Table~\ref{tab:powerlaw}), confirming that neither architecture introduces a measurable spectral bias on the D10 realization. The D10 low-latitude dataset yields the steepest power spectrum of all evaluated regions ($\alpha_d^{\rm input} = 3.151$), consistent with the greater concentration of dust power at large angular scales in the Galactic plane.

\begin{table}[H]
\centering
\caption{Power-law fit parameters of the angular power spectrum, $C_\ell = A_d\,\ell^{-\alpha_d}$, for the input maps and the outputs of the PINN and U-Net. The Model column indicates the reconstruction architecture, while Sky Region identifies the validation set or the corresponding dust realization (D4, D7 or D10) and Galactic latitude region. $A_d^{\rm input}$ and $\alpha_d^{\rm input}$ denote the amplitude and spectral index, respectively, obtained from the power-law fit to the input dust maps, while $A_d^{\rm output}$ and $\alpha_d^{\rm output}$ are the corresponding parameters obtained from the reconstructed maps. $\Delta\alpha_d$ [\%] gives the relative percentage difference between the input and output spectral indices, with smaller values indicating better preservation of the input spectral slope. All fitted parameters are reported as mean $\pm$ standard deviation over the corresponding dataset.}
\label{tab:powerlaw}
\resizebox{\linewidth}{!}{%
\begin{tabular}{llccccc}
\hline
\textbf{Model} & \textbf{Sky Region} & $A_d^{\rm input}$ & $\alpha_d^{\rm input}$ & $A_d^{\rm output}$ & $\alpha_d^{\rm output}$ & $\Delta\alpha_d$ [\%] \\
\hline
\multirow{7}{*}{PINN}
 & Validation   & $2.98\times10^{4} \pm 5.40\times10^{4}$ & $2.891 \pm 0.298$ & $3.12\times10^{4} \pm 5.62\times10^{4}$ & $2.891 \pm 0.296$ & $0.004$ \\
 & d4 high lat  & $5.02 \pm 1.04$                         & $2.632 \pm 0.034$ & $4.62 \pm 1.19$                         & $2.615 \pm 0.042$ & $0.653$ \\
 & d4 low lat   & $1.84\times10^{5} \pm 3.48\times10^{5}$ & $3.062 \pm 0.313$ & $2.09\times10^{5} \pm 3.95\times10^{5}$ & $3.078 \pm 0.312$ & $0.502$ \\
 & d7 high lat  & $7.94 \pm 0.89$                         & $2.685 \pm 0.018$ & $7.96 \pm 0.80$                         & $2.681 \pm 0.016$ & $0.138$ \\
 & d7 low lat   & $1.31\times10^{5} \pm 2.48\times10^{5}$ & $2.949 \pm 0.311$ & $1.47\times10^{5} \pm 2.78\times10^{5}$ & $2.962 \pm 0.310$ & $0.451$ \\
 & d10 high lat & $8.11 \pm 2.03$                         & $2.737 \pm 0.041$ & $7.45 \pm 2.17$                         & $2.721 \pm 0.048$ & $0.591$ \\
 & d10 low lat  & $3.03\times10^{5} \pm 5.97\times10^{5}$ & $3.151 \pm 0.326$ & $3.40\times10^{5} \pm 6.71\times10^{5}$ & $3.171 \pm 0.325$ & $0.637$ \\
\hline
\multirow{7}{*}{U-Net}
 & Validation   & $2.98\times10^{4} \pm 5.40\times10^{4}$ & $2.891 \pm 0.298$ & $3.03\times10^{4} \pm 5.30\times10^{4}$ & $2.894 \pm 0.287$ & $0.094$ \\
 & d4 high lat  & $5.02 \pm 1.04$                         & $2.632 \pm 0.034$ & $4.86 \pm 1.24$                         & $2.627 \pm 0.042$ & $0.190$ \\
 & d4 low lat   & $1.84\times10^{5} \pm 3.48\times10^{5}$ & $3.062 \pm 0.313$ & $1.93\times10^{5} \pm 3.59\times10^{5}$ & $3.074 \pm 0.306$ & $0.367$ \\
 & d7 high lat  & $7.94 \pm 0.89$                         & $2.685 \pm 0.018$ & $8.60 \pm 0.84$                         & $2.697 \pm 0.016$ & $0.450$ \\
 & d7 low lat   & $1.31\times10^{5} \pm 2.48\times10^{5}$ & $2.949 \pm 0.311$ & $1.41\times10^{5} \pm 2.60\times10^{5}$ & $2.962 \pm 0.303$ & $0.454$ \\
 & d10 high lat & $8.11 \pm 2.03$                         & $2.737 \pm 0.041$ & $7.98 \pm 2.48$                         & $2.729 \pm 0.051$ & $0.315$ \\
 & d10 low lat  & $3.03\times10^{5} \pm 5.97\times10^{5}$ & $3.151 \pm 0.326$ & $2.99\times10^{5} \pm 5.90\times10^{5}$ & $3.150 \pm 0.325$ & $0.035$ \\
\hline
\end{tabular}
}
\end{table}

\subsection{Validation with Other Dust Models}

To assess generalization beyond the training distribution, both architectures are evaluated on two additional Galactic dust realizations, D4 and D7, each split into high-latitude ($|b| > 30^\circ$) and low-latitude ($|b| < 30^\circ$) subsets of 1000 patches, following the same procedure used for D10 in Section~\ref{sec:validation_simulations}. Each test dataset is further combined with independent randomized noise realizations, ensuring that the evaluation captures generalization across both dust morphology and noise properties. Neither the dust realization nor the specific noise draw was seen during training, so these datasets provide an independent test of robustness to unseen Galactic dust morphology and instrumental noise statistics.

\begin{figure}[H]
\centering
\begin{minipage}{\linewidth}
\centering
\includegraphics[width=\linewidth]{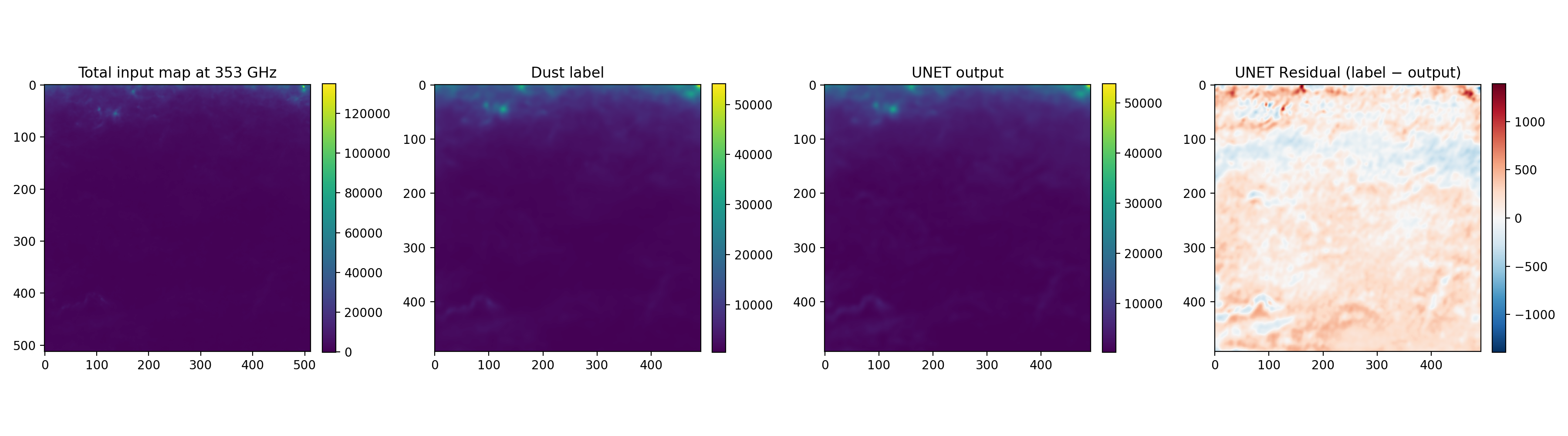}
\end{minipage}
\begin{minipage}{\linewidth}
\centering
\includegraphics[width=\linewidth]{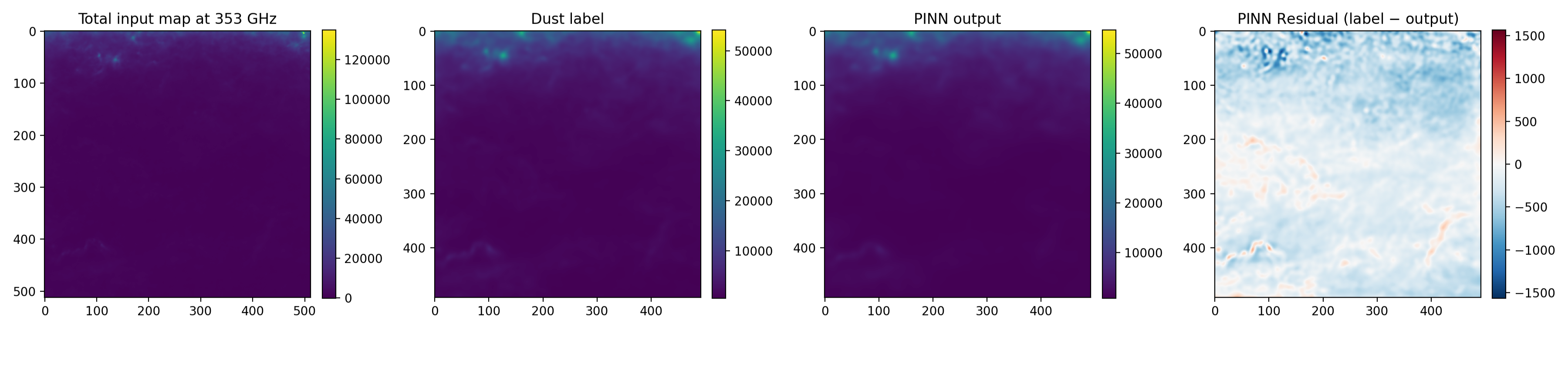}
\end{minipage}
\caption{An example of the reconstruction of a low-latitude patch ($|b| < 30^\circ$) from a sky with the D4 dust emission. Rows and columns represent the same as in Figure \ref{fig:cl_d10_low_patch}.}
\label{fig:cl_d4_low_patch}
\end{figure}

\begin{figure}[H]
\centering
\begin{minipage}{\linewidth}
\centering
\includegraphics[width=\linewidth]{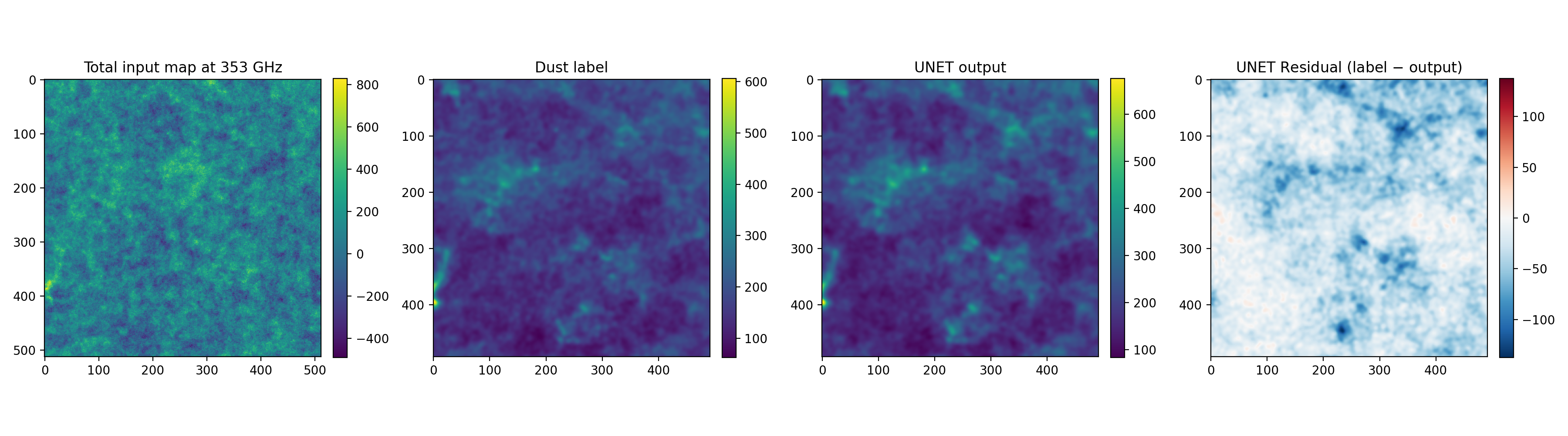}
\end{minipage}
\begin{minipage}{\linewidth}
\centering
\includegraphics[width=\linewidth]{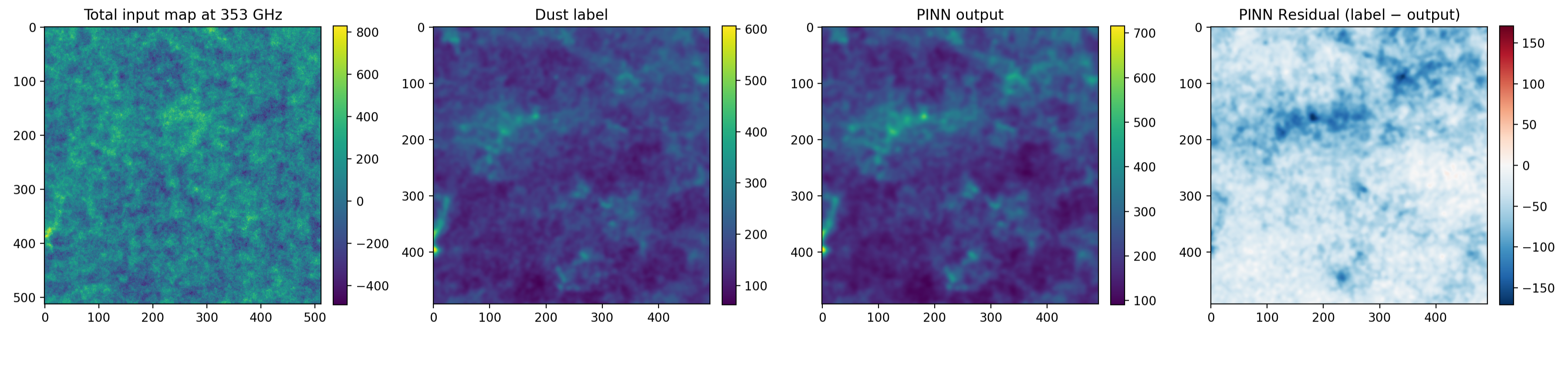}
\end{minipage}
\caption{An example of the reconstruction of a high-latitude patch ($|b| > 30^\circ$) from a sky with the D4 dust emission. Rows and columns represent the same as in Figure \ref{fig:cl_d10_low_patch}.}
\label{fig:cl_d4_high_patch}
\end{figure}

\begin{figure}[H]
\centering
\begin{minipage}{\linewidth}
\centering
\includegraphics[width=\linewidth]{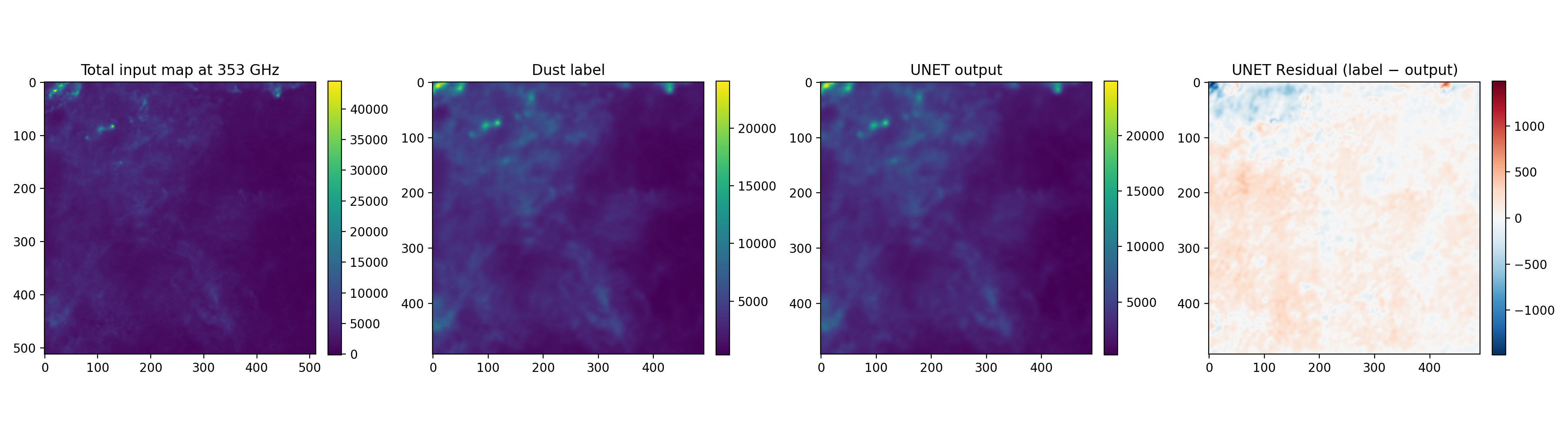}
\end{minipage}
\begin{minipage}{\linewidth}
\centering
\includegraphics[width=\linewidth]{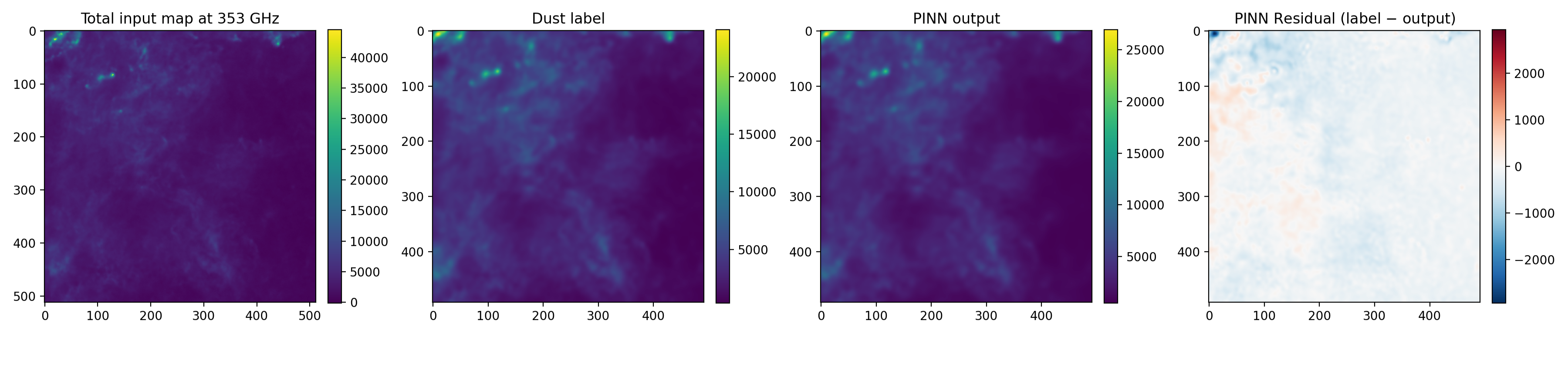}
\end{minipage}
\caption{An example of the reconstruction of a low-latitude patch ($|b| < 30^\circ$) from a sky with the D7 dust emission. Rows and columns represent the same as in Figure \ref{fig:cl_d10_low_patch}.}
\label{fig:cl_d7_low_patch}
\end{figure}

\begin{figure}[H]
\centering
\begin{minipage}{\linewidth}
\centering
\includegraphics[width=\linewidth]{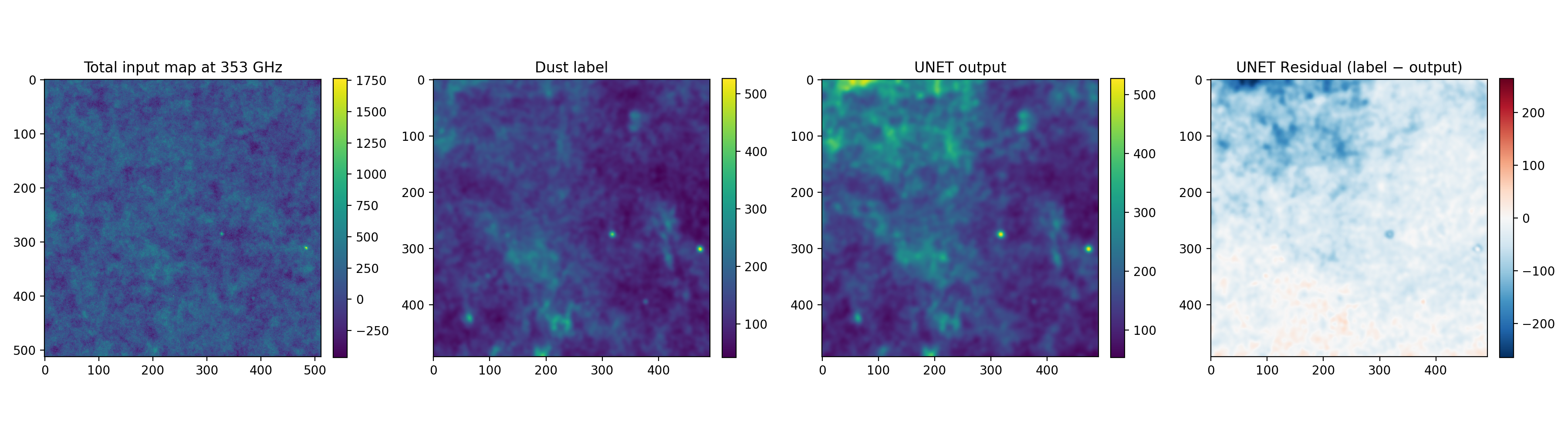}
\end{minipage}
\begin{minipage}{\linewidth}
\centering
\includegraphics[width=\linewidth]{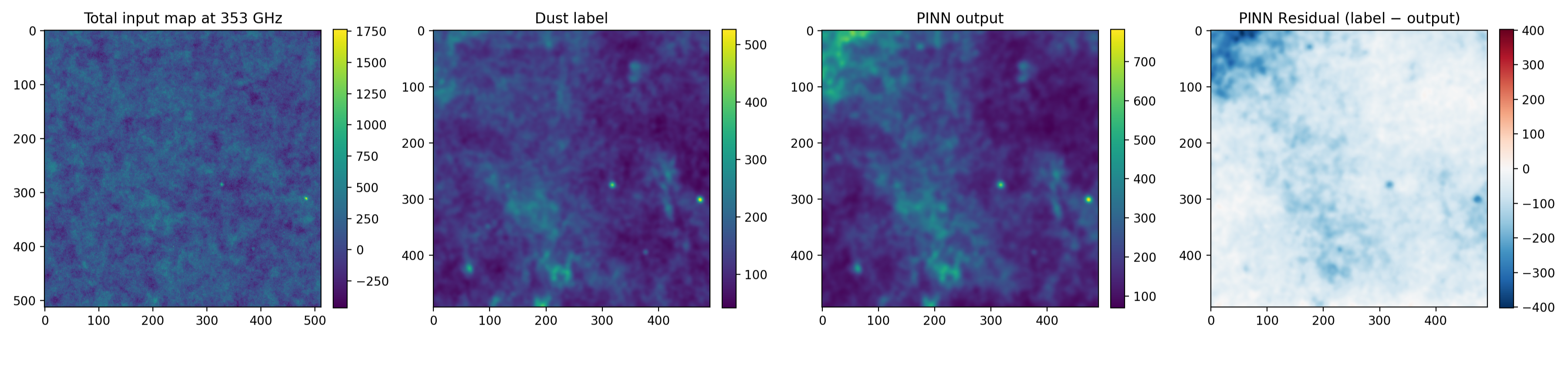}
\end{minipage}
\caption{An example of the reconstruction of a high-latitude patch ($|b| > 30^\circ$) from a sky with the D7 dust emission. Rows and columns represent the same as in Figure \ref{fig:cl_d10_low_patch}.}
\label{fig:cl_d7_high_patch}
\end{figure}

Representative patch reconstructions are shown in Figures~\ref{fig:cl_d4_low_patch}--\ref{fig:cl_d7_high_patch}. As shown, both networks recover the large-scale dust morphology in all four cases, with residuals visibly concentrated around bright point sources at high latitude, consistent with the D10 results discussed above.

The corresponding quantitative metrics (Table~\ref{tab:reconstruction}) follow the same qualitative trend observed for D10: the U-Net attains marginally lower RMSE and MAE than the PINN in three of the four D4/D7 subsets, with the largest relative gap at D4 low latitude (RMSE $4.38\times10^{-4}$ for the U-Net versus $5.33\times10^{-4}$ for the PINN, a $\sim$18\% difference). The exception is D7 high latitude, where the PINN attains a slightly higher SSIM ($0.972$ vs.\ $0.971$) despite a marginally larger RMSE, indicating that pixel-wise error and structural similarity do not always favor the same architecture. As in the D10 case, differences in PSNR between the two networks remain below 0.5\,dB across all four subsets, and SSIM values exceed 0.96 in every case, indicating that the physical constraint imposed by the MBB layer does not measurably degrade morphological fidelity even on dust realizations absent from training.

The power-law fits (Table~\ref{tab:powerlaw}) show that both architectures preserve the spectral index of the dust power spectrum to well within 1\% for D4 and D7, mirroring the D10 result. Notably, the ordering between architectures is not consistent across datasets: the PINN achieves the closest agreement with the input spectral index at D7 high latitude ($\Delta\alpha_d = 0.138\%$, versus $0.450\%$ for the U-Net), while the U-Net is closer at D4 high latitude ($\Delta\alpha_d = 0.190\%$ versus $0.653\%$ for the PINN). This absence of a systematic advantage for either architecture across the three independent dust realizations (D4, D7, D10) suggests that the spectral fidelity of the reconstruction is primarily determined by the training procedure and data statistics rather than by the presence or absence of the physical constraint layer.

Taken together, the D4 and D7 results corroborate the pattern established with the validation set and D10, the U-Net retains a small, consistent advantage in pixel-wise error metrics, attributable to its additional degree of freedom relative to the three-parameter physically constrained output of the PINN, while both architectures generalize comparably well to unseen dust morphology and noise realizations in terms of structural similarity and angular power spectrum preservation.

\section{Discussion and Conclusions}

This work presented a methodological comparison between a standard U-Net and a physics-informed neural network (PINN) for the reconstruction of Galactic thermal dust emission at 353\,GHz from realistic multi-frequency microwave simulations. The central architectural difference between the two models is that the PINN embeds the Modified Black Body (MBB) spectral law as a hard, differentiable constraint at the output layer, whereas the U-Net produces the target map directly with no explicit physical parameterization.

Across the validation set and the three independent dust realizations used for generalization testing (D4, D7, D10), both architectures achieve comparable reconstruction quality, with SSIM consistently above 0.96 and PSNR in a similar range for both models in every evaluated region, and both preserve the angular power-spectrum spectral index to within 1\% of the input in every tested region. The U-Net attains marginally lower RMSE and MAE in the majority of the tested datasets, which is consistent with it having one fewer layer of representational constraint: its single-channel output is optimized directly against the target map, whereas the PINN's output is mediated by three intermediate parameters combined through a fixed, non-trainable physical equation. That the PINN nonetheless matches the U-Net in structural similarity and in the preservation of the angular power spectrum shape indicates that the additional constraint does not come at a significant cost to reconstruction fidelity, which is the primary methodological question this comparison was designed to address.

The generalization tests on D4, D7 and D10 further show that neither architecture exhibits a systematic degradation in performance relative to the validation set, and that the relative ranking between the two models (in RMSE, MAE, SSIM, PSNR and fitted spectral index $\alpha_d$) is not consistent across sky regions or dust models. This suggests that, at the level of aggregate reconstruction metrics, the choice between the two architectures does not have a uniform effect on generalization capability under the conditions tested here, and that any advantage of one model over the other is likely to be region- and dataset-dependent rather than a general property of the physics-informed constraint.

The main methodological contribution of the PINN formulation is therefore not a gain in reconstruction accuracy but a difference in the structure of the output: every PINN prediction corresponds by construction to a thermodynamically admissible combination of emission amplitude, spectral index and temperature, whereas the U-Net offers no equivalent guarantee or intermediate representation. Evaluating whether these intermediate parameters carry physically meaningful information, and how they compare to independent physical measurements, falls outside the scope of the present comparison and is left for dedicated analysis.

A limitation of the current setup is that both networks are optimized against a single target frequency (353\,GHz), which constrains the class of problems this comparison speaks to: a single-frequency reconstruction task in which the physical constraint acts primarily as an architectural regularizer rather than as a mechanism for resolving a multi-observable inverse problem. Extending the training objective to a multi-frequency consistency loss, in which the network is required to reproduce the dust emission simultaneously at all available frequencies, would test the physics-informed formulation under conditions closer to a genuinely overdetermined inverse problem, and constitutes a natural next step for this line of work. Validation against real observational data, which would assess the transferability of both trained models beyond the Planck Sky Model simulations used here, is likewise left for future work.

In summary, embedding the MBB spectral model as a hard architectural constraint in the PINN does not come at the cost of reconstruction accuracy relative to the unconstrained U-Net baseline, while additionally guaranteeing that every PINN output corresponds to a thermodynamically admissible physical configuration. This establishes the physics-informed formulation as a viable methodological alternative for dust component separation, offering physical interpretability without a trade-off in reconstruction fidelity.

\funding{N.M.J, J.M.C and L.B acknowledge the CNS2022-135748 / MCIN/AEI/10.13039/501100011033 project.}

\roles{Conceptualization, J.M.C. and L.B.; methodology, N.M.J., J.M.C. and L.B.; software, N.M.J.; validation, N.M.J., J.M.C. and L.B.; formal analysis, N.M.J.; investigation, N.M.J.; resources, J.M.C. and L.B.; data curation, N.M.J.; writing---original draft preparation, N.M.J.; writing---review and editing, J.M.C. and L.B.; visualization, N.M.J.; supervision, J.M.C. and L.B.; project administration, J.M.C. and L.B.; funding acquisition, J.M.C. and L.B.}

\section*{Conflict of interest}
The authors declare no conflicts of interest.

\bibliographystyle{unsrt}
\bibliography{references}

\end{document}